\documentclass{ijuc}
\usepackage[pdftex]{graphics}

\usepackage[T1]{fontenc}
\usepackage[english]{babel}
\usepackage[utf8]{inputenc}
\usepackage[T1]{fontenc}
\usepackage{microtype}
\usepackage[dvips]{graphicx}
\usepackage{xcolor}
\usepackage{times}
\usepackage{pdfpages}
\usepackage{appendix}
\usepackage{titlesec}
\usepackage{lscape}
\usepackage{amsmath}
\usepackage{amsthm}
\usepackage{amssymb}
\usepackage{amsfonts}
\usepackage{cite}
\usepackage{braket}
\usepackage{comment}
\usepackage{algorithm}
\usepackage{algpseudocode}
\usepackage{hyperref}
\usepackage{color}

\usepackage{hyperref}   

\begin{document}

\title{Optimizing quantum circuits with evolutionary algorithms for stable Boolean gates, elementary cellular automata, and highly entangled quantum states}

\author{Shailendra Bhandari\inst{1,2}\email{shailendra.bhandari@oslomet.no} 
 \and
  Stefano Nichele\inst{1,2,4} 
\and
  Sergiy Denysov\inst{1}
\and
  Pedro G. Lind\inst{1,2,3,5} 
}

\institute{Department of Computer Science, OsloMet -- Oslo Metropolitan University, P.O.~Box 4 St.~Olavs plass, N-0130 Oslo, Norway \and
OsloMet Artificial Intelligence Lab, Pilestredet 52, N-0166 Oslo, Norway \and
Kristiania University of Applied Sciences, Oslo, Norway \and
Department of Computer Science and Communication, Østfold University College, B R A Veien 4, N-1757 Halden, Norway \and
Simula Research Laboratory, Numerical Analysis and Scientific Computing, Oslo, 0164, Norway}


\maketitle            

\begin{abstract}
We investigate the potential of bio-inspired evolutionary algorithms (EAs) for designing quantum circuits with specific goals, focusing on two particular tasks. The first task involves using these algorithms to reproduce stochastic cellular automata with given rules. We test the robustness of quantum implementations of the cellular automata for different numbers of quantum gates. The second task deals with sampling quantum circuits that generate highly entangled quantum states, which constitute an important resource for quantum computing.  In particular, an evolutionary algorithm is employed to optimize circuits with respect to a fitness function defined with the Meyer-Wallach (MW) entanglement measure. We demonstrate that, by balancing the mutation rate between exploration and exploitation, we can find entangling quantum circuits for up to five qubits. We also discuss the trade-off between the number of gates in quantum circuits and the computational costs of finding the gate arrangements leading to a strongly entangled state.  Our findings provide additional insight into the trade-off between the complexity of a circuit and its performance, which is an important factor in the design of quantum circuits.
  
\end{abstract}


\section{Introduction}

Quantum computing utilizes principles of quantum mechanics to perform certain computations faster and on larger scales than classical computers are capable of~\cite{nielsen_chuang_2010}. 
Quantum bits, or {qubits}, are key building blocks of quantum computers that enable the latter to be advantageous in solving complex problems~ \cite{nielsen_chuang_2010}. 
The intersection of quantum computing and artificial intelligence (AI) is currently an area of active research that has the potential to revolutionize various fields and industry sectors. 
Quantum Machine Learning (QML) algorithms, particularly EAs, belong to one of the research directions actively explored now~\cite{Biamonte2017, 2014, Tacchino2019}. It is expected that the new results will help to speed up the optimization process further and improve the overall performance of computational algorithms. In particular, Evolutionary Quantum Computation (EQC) is an emerging approach that blends EAs and quantum computing to solve complex optimization problems~\cite{https://doi.org/10.48550/arxiv.2302.01303, Brown_2023, ballinas2022hybrid}, with recent applications including quantum circuit optimization \cite{sunkel2023ga4qcogeneticalgorithmquantum}, quantum state engineering \cite{Brown_2023}, and hybrid quantum genetic algorithms for optimization tasks \cite{Ballinas2022}. It is expected that these advancements will help to speed up the optimization process and improve the overall performance of computational algorithms.

The aim of this work is twofold. First, we would like to show how quantum algorithms can be used to solve specific problems related to the engineering of EAs. Next, in a complementary manner, we want to illustrate how EAs can be used to design quantum circuits fulfilling specific requirements. In particular, we will address the problem of designing quantum circuits that process the initial product states into
highly entangled states.

{Entanglement}, one of the key concepts of quantum physics, also plays a crucial role in quantum information processing \cite{articleQE}. The term `entangled state' refers to a state of several quantum objects that are intrinsically correlated with each other, in a way different from all possible classical correlations~\cite{articleQuantumEnranglement}. For example,  when two qubits are entangled, changes in the state of one qubit affect the state of the other one. There are several measures proposed to quantify the strength of these essentially quantum correlations, and so a quantitative definition of the word `highly,' strictly speaking, depends on the chosen measure. Highly entangled quantum systems are often difficult to simulate on classical computers\cite{PhysRevLett.91.147902,2008}, though certain exceptions, such as states generated by Clifford circuits, can be efficiently simulated\cite{Addala2025}. This complexity has significant implications for quantum information processing technologies. Metaphorically speaking, highly entangled qubit states serve as fuel to quantum information processing, and without them, the latter loses its superiority with respect to classical processing~\cite{PhysRevA.100.022342}.
To create a highly entangled state out of a non-entangled one is not an easy task. And even if we have managed to prepare a maximally entangled state, it can become less entangled just because the quantum system prepared in this state is interacting with the environment~\cite{Protocols, ACAMPORA2021542,Amal2022}. All real quantum systems suffer from this interaction, whose action is conventionally described with the term ``decoherence''. Therefore, it is important to keep the time during which decoherence strikes the system as short as possible. Practically, this means minimizing the computation time. In addition, every time a quantum gate (that is, a transformation of the state of a system of qubits) is implemented, an additional error occurs due to the non-ideal character of the implementation. After passing through many gates, the state
is no longer reliable in the sense that it is very different from what it has to be according to the corresponding quantum program. Thus, the more complicated the quantum circuit is (in terms of the number of gates it consists of), the stronger the joint action of decoherence and accumulated implementation errors~\cite{Dancing}. As has been mentioned above, not all entangled states are equal in their `strength'. This means that we need to quantify and measure the amount of entanglement between different qubits. This brings another challenge, because, in a generic quantum system, the quantification of entanglement is a non-trivial task~ \cite{friis2019entanglement}.

In Section \ref{sec:methods} we outline the methodology and introduce the key concepts.  
Section \ref{results} then presents two main results.  
Subsection \ref{sec:sca} shows how classical stochastic cellular automata can be implemented with quantum circuits, extending earlier work on complex-system modeling \cite{10.1007/978-3-031-14926-9_11}.  
Subsection \ref{sec:entangle} introduces a novel use of evolutionary algorithms to design circuits that maximize multi-qubit entanglement, evaluating each candidate with the Meyer–Wallach measure.  
Finally, Section \ref{sec:conclusions} summarizes the findings and outlines future directions.

\section{Methods}
\label{sec:methods}

Our study implements two distinct methods for designing quantum circuits. The first method utilizes cellular automata rules combined with the KL fitness function. Here we advanced the original framework from Ref. \cite{10.1007/978-3-031-14926-9_11} for generating quantum circuits that replicate quantum circuits with EAs. Upon this groundwork, the second method focuses on designing quantum circuits for maximum entanglement, using the MW measure and von Neumann entropy for entanglement evaluation. The comprehensive details regarding the implementation of evolutionary quantum algorithms for both methodologies are thoroughly delineated in \textbf{Supplementary Note I}.

\subsection{Evolutionary algorithms: Mutations and fitness function} 
\label{FF}
The evolutionary algorithm initiates with a specific assigned number of chromosomes, each populated with a set number of quantum gates within the circuit. This initial population is then subject to evolution, where the fittest chromosomes are identified, iteratively refining the existing ones, and introducing the best-fit chromosome at the forefront of each new generation. This evolutionary cycle continued for a predetermined number of generations, with each generation's best chromosome becoming the parent for the next. Throughout this process, the chromosomes are subjected to mutations to enhance optimization. The mutation process is introduced in two different forms: either by swapping existing gates in the chromosome with randomly selected ones from the gate pool, or by entirely replacing the chromosome to identify the best four parental candidates suitable for the next generation. The mutation process is controlled by a set of probabilities. The four best chromosomes are preserved as "elites" without changes, whereas the remaining chromosomes are selected for evolution based on their fitness, with selection changes directly linked to their performance.

In this context, we considered one-dimensional CAs where each of the cells can be in one of two possible states, 0 or 1 (Boolean CA), The evolution of these CA is governed by a set of rules, precisely 256 possible ones, that determine how the state of a cell updates at each step based on the states of its immediate neighbors. The rules correspond to each of the eight possible configurations of a cell and its two immediate neighbors, namely [0, 0, 0], [0, 0, 1], [0, 1, 0], [0, 1, 1], [1, 0, 0], [1, 0, 1], [1, 1, 0], and [1, 1, 1]. The rule dictates the next state of the central cell in these configurations. Table~\ref{tab:probabilities_exp3} shows the matching in all cases considered in this paper. 

To evaluate the performance of the quantum circuits, we measure only the first qubit (q0) and consider the fitness function as the $D_{KL}$, which measures the difference between two probability distributions and has been used in other works \cite{Martin2015, Lucas2019} as a fitness function:
\begin{equation}\label{eq:FF2}
D_{KL}(P||Q)= \sum_{\omega \in \Omega} P(\omega)log\Big(\frac{P(\omega)}{Q(\omega)}\Big).
\end{equation}

In our case, the distributions are discrete, and each one has eight different values related to the eight different initial states of the three qubits. If the distributions are a perfect match, then the fitness function is zero. 
\subsection{Assessing entanglement in basic quantum states of qubits}

Consider, for the start, the state of a system of two-qubit configurations of a quantum system. The state $\ket{\psi}$ is a superposition of the basis states, with $\alpha_{ij}$ representing the complex amplitudes for each basis state $\ket{ij}$, where $i,j\in\{0,1\}$. The most general form of a state for such a system can be written as: 
\begin{equation} 
\label{5}
\begin{split}
|\psi\rangle = \alpha_{00}\ket{00}+\alpha_{01}\ket{01}+\alpha_{10}\ket{10}+\alpha_{11}\ket{11}\,\, ,
\end{split}
\end{equation}
where $\alpha_{00}$, $\alpha_{01}$, $\alpha_{10}$, and $\alpha_{11}$ are complex amplitudes of the corresponding basis states. The normalization of the basis state is
\begin{equation}
    {|\alpha_{00}|}^2 + {|\alpha_{01}|}^2 + {|\alpha_{10}|}^2 + {|\alpha_{11}|}^2 
    = \sum_{X\in\{0,1\}^2}{|\alpha_X|}^2 = 1\, ,
\end{equation}
where $X\in\{0,1\}^2$ means all binary combinations of length 2.
For a two-qubit state, there is one fully entangled state given by \cite{PhysRevLett.91.147902,doi:10.1080/14789940801912366}
\begin{equation} \label{3}
|\psi\rangle = \frac{1}{\sqrt{2}} \left( \ket{00} + \ket{11} \right) \, .
\end{equation}
This state exemplifies quantum entanglement where the measurement outcome of one qubit directly determines the outcome of the other, showcasing the non-classical correlation inherent in quantum systems \cite{nielsen_chuang_2010}.
To contrast, consider the following non-entangled state:
\begin{equation}
    \ket{\psi} = \frac{1}{2}\ket{00}+\frac{1}{2}\ket{01}+\frac{1}{2}\ket{10}+\frac{1}{2}\ket{11}.
\end{equation}
In this state, when measuring one qubit, there is no certainty about the state of the other qubit. Namely, there is a probability of $1/2$ that the other qubit is in state $0$ and a probability of $1/2$ that it is in state $1$.

In general, suppose we have two quantum systems \( Q_1 \) and \( Q_2 \). These systems are said to be entangled if their joint quantum state cannot be factored into a product of individual states. More formally, a two-qubit state \( |\psi\rangle \) is entangled if it cannot be written as the tensor product of two pure states, i.e., $
|\psi\rangle \neq |\psi_1\rangle \otimes |\psi_2\rangle$ , where \( |\psi_1\rangle \) and \( |\psi_2\rangle \) are states belonging to individual systems \( Q_1 \) and \( Q_2 \), respectively. This condition holds for pure states, and for mixed states, the density matrix of the system cannot be written as a convex combination of product states. The simplest examples of entangled states are the Bell states, which represent maximally entangled two-qubit states \cite{HWeinfurter_1994}, and they enable tasks such as quantum cryptography \cite{RevModPhys.74.145}, superdense coding \cite{PhysRevLett.92.187901}, teleportation \cite{PhysRevLett.70.1895}, and entanglement swapping \cite{Mooney2019}. These Bell states are:
\[
\begin{aligned}
|\Phi^+\rangle &= \frac{1}{\sqrt{2}} \left( |0\rangle \otimes |0\rangle + |1\rangle \otimes |1\rangle \right), \quad
|\Phi^-\rangle = \frac{1}{\sqrt{2}} \left( |0\rangle \otimes |0\rangle - |1\rangle \otimes |1\rangle \right), \\
|\Psi^+\rangle &= \frac{1}{\sqrt{2}} \left( |0\rangle \otimes |1\rangle + |1\rangle \otimes |0\rangle \right), \quad
|\Psi^-\rangle = \frac{1}{\sqrt{2}} \left( |0\rangle \otimes |1\rangle - |1\rangle \otimes |0\rangle \right)
\end{aligned}
\] where \( |00\rangle, |01\rangle, |10\rangle, |11\rangle \) are the computational basis states.


The descriptors, expressibility, and entangling capability have been used to study the capabilities of the parametrized quantum circuit by quantifying its deviation from random circuits to approach the research question of how much generalization is effective enough in a quantum circuit for a given task \cite{Sim_20191}. The expressibility of a quantum circuit is the ability to generate pure states that are well representative of the Hilbert space \cite{PhysRevA.71.032313}. In a single qubit, the expressibility corresponds to the circuit's ability to explore the Bloch sphere. Sim et al. \cite{Sim_20191} propose to quantify the ability of the quantum circuit to generate a pure state as a representative of Hilbert space by comparing the true distribution of the fidelities corresponding to the parameterized quantum circuit (PQC),  to the distribution of fidelities from the ensemble of Haar random states. 

For example, the measure of expressibility can be calculated by taking the $D_{KL}$ between the estimated fidelity distribution and that of the Haar-distributed ensemble \cite{PhysRevA.71.032313}. 
In this paper, we considered two measures, the MW \cite{doi:10.1063/1.1497700} and the entanglement entropy (EE)\cite{EisertJ2010Alft, nielsen_quantum_2002, Longo_2020}.
The MW entanglement measure is popular because of its scalability and ease of computation.
It considers a pure state of $n$ qubits of the form \cite{brennen2003observable}:
\begin{equation}\label{MW2}
    Q(\ket{\psi}) = \frac{4}{n}\sum_{j=1}^n D(\ket{\hat{u}^k}, \ket{\hat{v}^k})\,,
\end{equation}
where $\ket{\hat{u}^k}$ and $\ket{\hat{v}^k}$ denote non-normalized vectors belonging to the complex vector space $\mathbf{C^{2n-2}}$, where $n$ denotes the number of qubits in the quantum system. These vectors are obtained by projecting the pure state $\ket{\psi}$ of the $n$-qubit system onto the local basis states of the $k^{th}$ qubit, where $k \in {1, 2, ..., n}$. The symbol "$\;\hat{}\;$" above the vectors signifies their non-normalized status.
The function $ D(\ket{\hat{u}^k}, \ket{\hat{v}^k})$ measures a distance between the two vectors $\ket{\hat{u}^k}$ and $ \ket{\hat{v}^k}$, measured through the generalized cross-product: 
\begin{equation}
    D(\ket{\hat{u}^k}, \ket{\hat{v}^k}) = \sum_{i<j}|{\hat{u}_i^k\hat{v}_j^k-\hat{u}_j^k\hat{v}_i^k}|^2 \, .
\end{equation}

Since the purity of the state of qubit $k$ is given by $Tr[\rho_k^2]=\braket{\hat{x}^k|\hat{x}^k}^2+\braket{\hat{y}^k|\hat{y}^k}^2$, we obtain \cite{brennen2003observable}  

\begin{equation}
     D(\ket{\hat{x}^k}, \ket{\hat{y}^k}) = \sum_{i<j}|{\hat{u}_i^k\hat{v}_j^k-\hat{u}_j^k\hat{v}_i^k}|^2
\end{equation}
It is because generalized cross product under logical unitaries, $D(\ket{\hat{u}^k}, \ket{\hat{v}^k})= D(\ket{\hat{x}^k}, \ket{\hat{y}^k})$ in relation to the norm of an anti-symmetric tensor $M_k = \ket{\hat{x}^k}\bra{\hat{y}^{*k}} - \ket{\hat{y}^k}\bra{\hat{x}^{*k}}$. Therefore, one arrives at an entanglement measure given by:
\begin{equation} \label{MW_equation}
    Q(\ket{\psi})= 2 \left ( 
                     1-\frac{1}{n}\sum_{k=0}^{n-1}Tr[\rho_k^2]
                     \right ),
\end{equation}
which is the usual definition of the MW measure. A detailed derivation and discussion of this measure, especially its implications for both pure and mixed states, are provided in \textbf{Supplementary Note II}.

The entanglement entropy quantifies the quantum entanglement among two subsystems within a larger quantum system. Specifically, for a pure bipartite quantum state of the mixed system, it is feasible to derive a reduced density matrix that characterizes the state of one of the subsystems \cite{nielsen_quantum_2002}. The entropy of entanglement is defined as the von Neumann entropy of the reduced density matrix for either of the subsystems \cite{EisertJ2010Alft}.
The von Neumann entropy quantifies the uncertainty or randomness in a quantum state and measures the degree of entanglement between the subsystems \cite{nielsen_quantum_2002}. If the entropy of entanglement is non-zero, it implies that the subsystem is in a mixed state, and consequently, the two subsystems are entangled \cite{EisertJ2010Alft}. Therefore, the entropy of entanglement offers a useful method for quantifying the degree of entanglement between the subsystems of a composite quantum system.

The von Neumann entropy of entanglement is mathematically defined as the von Neumann entropy of the reduced density matrix for a subsystem within a composite quantum system \cite{Longo_2020}, namely
\begin{equation}\label{Von_neumann}
S(\rho_A) = -\mathrm{Tr}(\rho_A \log_2 \rho_A) \, ,
\end{equation}
where $\mathrm{Tr}$ represents the trace operation, and $\log_2$ denotes the base-2 logarithm. The von Neumann entropy of entanglement provides a measure of the uncertainty or randomness in the state of subsystem $A$. Thus, it serves as an indicator of the degree of entanglement between the subsystems \cite{PhysRevA.54.3824}. When dealing with a composite quantum system that's in a mixed state, the von Neumann entropy of entanglement measures the least amount of entanglement present across all possible pure-state decompositions of the mixed state. This method allows for measuring entanglement in a mixed state by considering the representation of the composite system that has minimal entanglement.

In what concerns implementation, we have used the mathematical formulations presented in Equations \eqref{MW_equation} and \eqref{Von_neumann} in a framework previously developed, aligned with the methodologies outlined in Ref. \cite{10.1007/978-3-031-14926-9_11}. The EAs have emerged as a promising tool for quantum circuit design \cite{sunkel2023ga4qco, Brown_2023, ballinas2022hybrid}. So our implementation of EAs for quantum circuit design aims to maximize entanglement using two specified methods. The framework employs an evolutionary algorithm to automatically generate quantum circuits that satisfy defined properties specified through a fitness function. The evolution properties can be controlled by tuning parameters such as initial population, number of generations, probability of mutation operator, number of gates in the circuit, and the chosen fitness function.  Details of the implementation process are available in \textbf{Supplementary Note III}.

\section{Results}\label{results}
\subsection{Realizing stochastic cellular automata with quantum circuits}
\label{sec:sca}


In this part, we use an evolutionary algorithm to realize cellular automata (CA) for specific rules, deterministic and stochastic.
The CAs are realized with quantum 
circuits and measurements. The circuits are constructed by assembling gates from a fixed set of five gates.  Our approach involves a mutation-based evolutionary algorithm, enabling the optimization of gate types and their placements.

Initially, a predetermined number of chromosomes is generated, corresponding to the number of quantum gates per circuit. Through iterative refinement, the genetic algorithm evolves the existing chromosomes into new variations, selecting the fittest chromosome within each generation to serve as the parent for the subsequent generation. The assessment of the performance of the derived quantum circuits revolves around the measurement of the first qubit,  $q_0$. In our quantum circuits, we denote qubits as $q_i$, where $i$ indexes the qubit position (e.g., $q_0$ is the first qubit in a three-qubit system). This assessment entails comparing the measured probability of an initial state, computed from the population, with the corresponding target probability. To quantify the difference between two sets of probability distributions,  we use the Kullback-Leibler divergence ($D_{KL}$) \cite{10.1348/000711010X522227} as a fitness function. The $D_{KL}$ fitness score measures the difference between two probability distributions, with a detailed formulation provided in Section \ref{FF}, (Equation \ref{eq:FF2}). In our case, these distributions are discrete, each corresponding to the eight different initial states of the three qubits. If the distributions completely match, the fitness function value is zero. A decreasing fitness value close to zero indicates an increase in the efficiency of the circuits. The resulting quantum circuits are evaluated through measurements, thereby demonstrating the algorithm's efficacy. For more details, see \textbf{Supplementary Note I}.%
    \begin{table}[t]
    \centering
    \small{
    \begin{tabular}{|l |l | l |l |l |l |l|}
        \hline
        Neighbors & Sto.~CAProb.
        & Rule90 & Rule110 & Prob.~1 & Prob.~2 &Prob.~3 \\[1ex] 
        \hline
        [0, 0, 0] & 0.394221 &0 & 0 & 0.6364 & 0.4778 & 0.1988\\
        \hline
        [0, 0, 1] & 0.094721 & 1 & 1 & 0.6603 & 0.5604 & 0.4701\\
        \hline
        [0, 1, 0] & 0.239492 & 0 & 1 & 0.5261 & 0.8528 & 0.9836\\
        \hline
        [0, 1, 1] & 0.408455 & 1 & 1 & 0.1748 & 0.4818 & 0.7115\\
        \hline
        [1, 0, 0] & 0 & 1 & 0 & 0.8820 & 0.3143 & 0.6616\\
        \hline
        [1, 0, 1] & 0.730203 &  0 & 1 & 0.3371 & 0.3464 & 0.1218\\
        \hline
        [1, 1, 0] & 0.915034 & 1 & 1 & 0.0340 & 0.0678 & 0.1328\\
        \hline
        [1, 1, 1] & 1 & 0 & 0 & 0.4444 & 0.9124 & 0.7306\\ 
        \hline
    \end{tabular}}
    \caption{%
      \label{tab:probabilities_exp3}
      The deterministic and stochastic CAs are considered in this paper. For stochastic cases, the values indicate the probability of an update of value 1 for the middle cells in the neighborhood of the triad.
      For the deterministic cases, the values indicate the exact update imposed.}
\end{table}
 
\begin{figure}[t]
    \centering
    \includegraphics[width=0.5\textwidth]{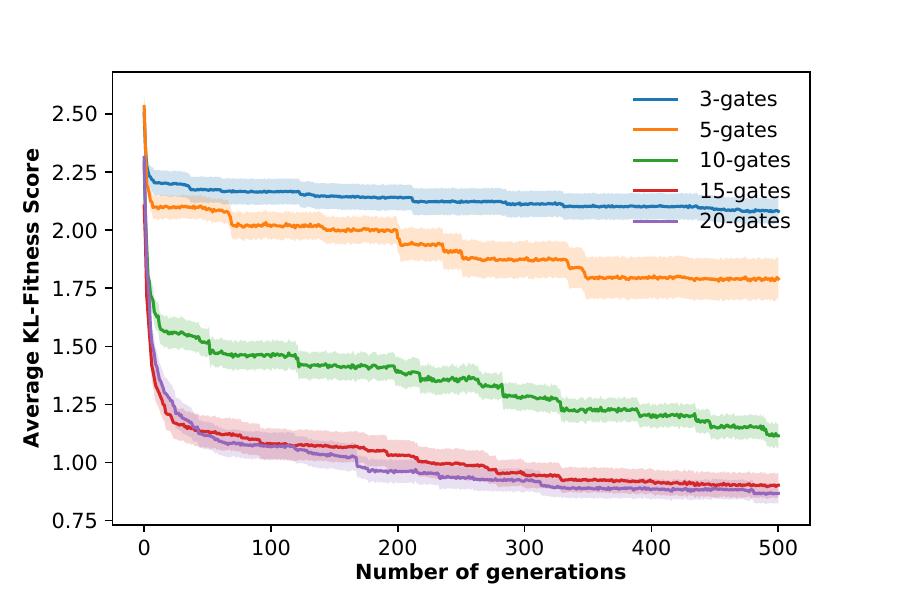}%
    \put(-150,0){\textbf{a)}} 
    \hfill
    \includegraphics[width=0.5\textwidth]{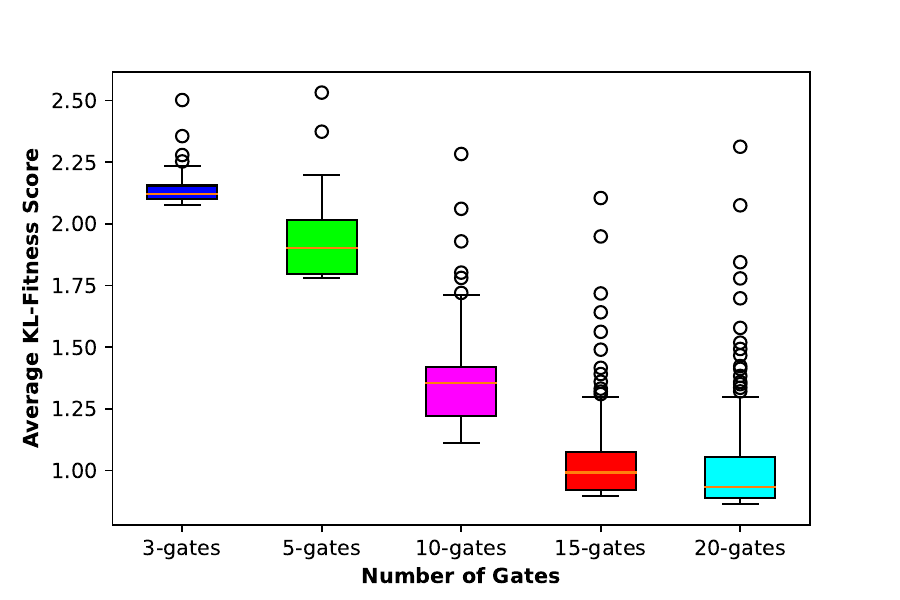}%
    \put(-150,0){\textbf{b)}} 
    \caption{\protect
       \textbf{a)} The fitness scores as a function of the number of generations, for {\bf different number of gates}.
       \textbf{b)} The number of gates vs. the best fitness scores. 
       The fitness scores of each gate for the box plots are the best fitness scores per run, and the fitness scores for the lower two plots are the average fitness scores of 50 runs.}
    \label{box_plot}
\end{figure}

We consider three different types of CAs, starting with a stochastic {\it critical} cellular automaton, followed by more general stochastic CA (random updates) and a few deterministic rules, namely, rule $90$ and $110$~ \cite{WolframStephen1984Caam}. The updates for each type of CA are shown in Table~\ref{tab:probabilities_exp3}. Our evolutionary algorithm uses a population of \( N_c \) chromosomes (where \( N_c \) is the number of candidate circuits), evolves over \( N_g \) generations (where \( N_g \) is the number of iterations), and tests \( N_{ic} \) initial conditions (where \( N_{ic} \) is the number of random starting states). In this study, we set \( N_c = 20 \), \( N_g = 500 \), and \( N_{ic} = 50 \) to enhance robustness, building on prior settings of \( N_c = 20 \), \( N_g = 150 \), and \( N_{ic} = 20 \)~\cite{10.1007/978-3-031-14926-9_11}. Recent advancements in quantum cellular automata (QCA) models further contextualize our work. For instance, Arrighi et al. (2022) \cite{arriola2020bio} demonstrated that QCA can simulate complex lattice gauge theories, suggesting a quantum analog to our classical CA replication, while Jones et al. (2022) \cite{jones2022small} experimentally realized a QCA on a 23-qubit digital superconducting processor, achieving robust rule replication under noise, paralleling our focus on stability across gate counts (Figure \ref{box_plot}).

\begin{figure}[t]
    \centering
    \includegraphics[width=0.5\textwidth]{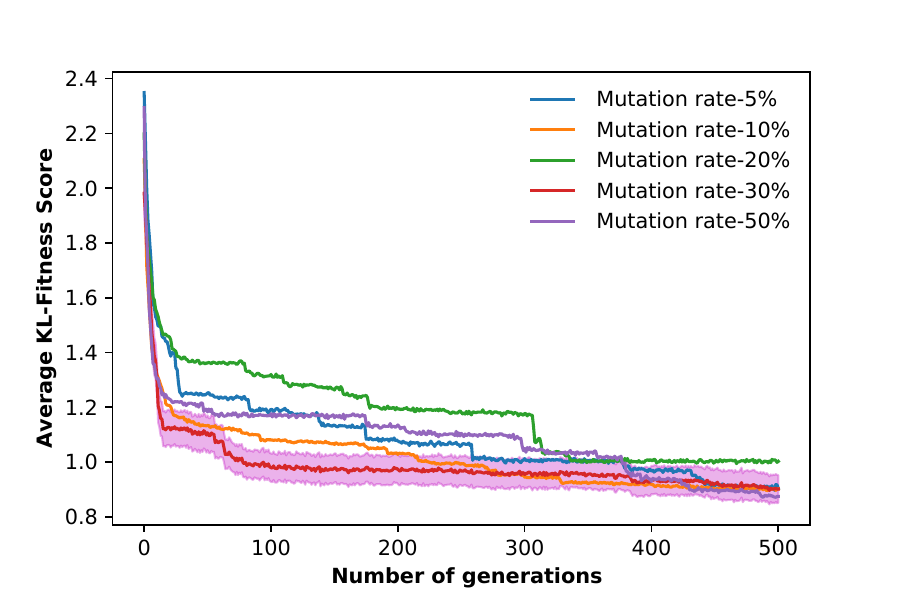}%
    \put(-150,0){\textbf{a)}} 
    \hfill
    \includegraphics[width=0.5\textwidth]{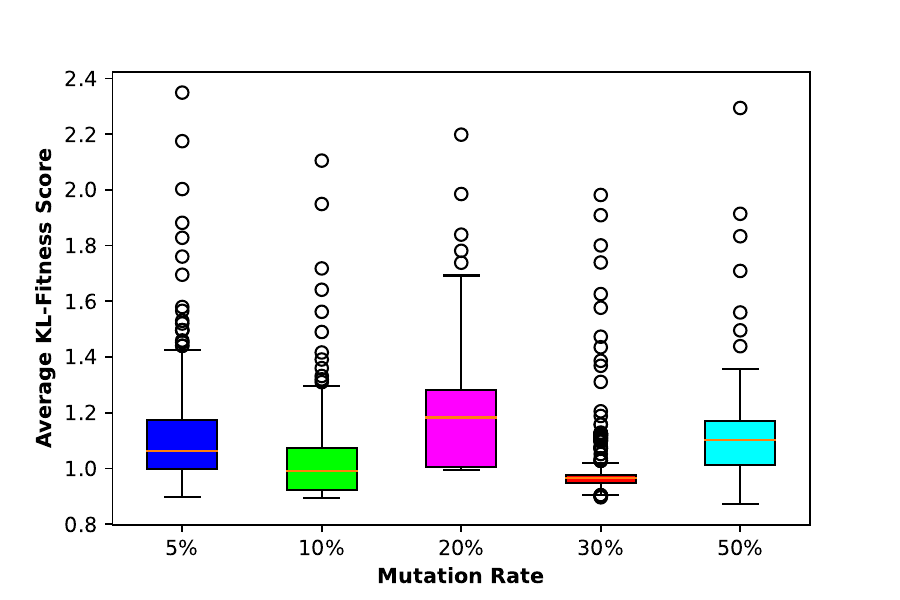}%
    \put(-150,0){\textbf{b)}} 
    \caption{\protect
       \textbf{a)} The fitness scores as a function of the number of generations, for {\bf different mutation rates}.
       \textbf{b)} Mutation rates vs. the best fitness scores. 
       The fitness scores of each gate are the average fitness scores of 50 runs.}
    \label{img:mutation_rate}
\end{figure}

\begin{figure}[t]
    \centering
    \includegraphics[width=0.495\textwidth]{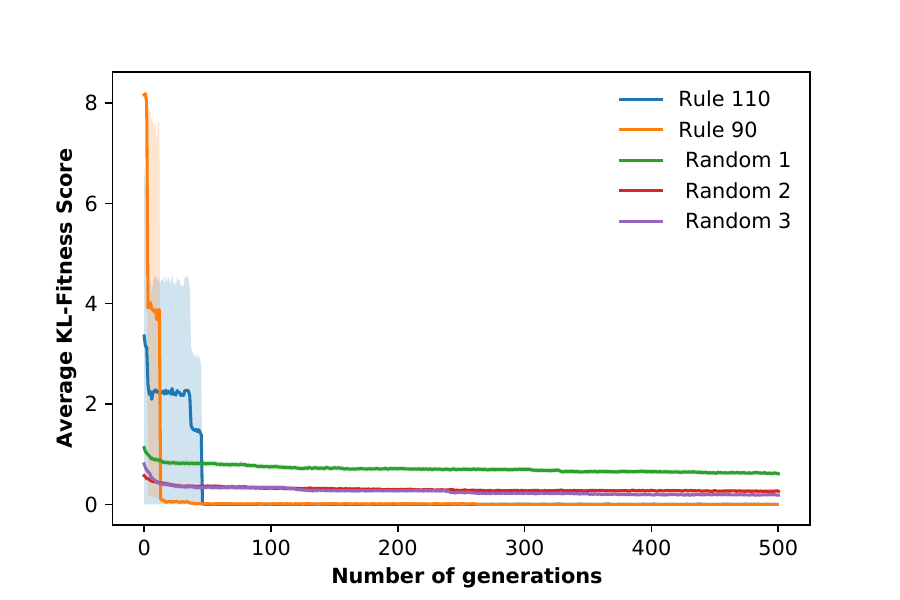}%
    \put(-150,0){\textbf{a)}} 
    \hfill
    \includegraphics[width=0.495\textwidth]{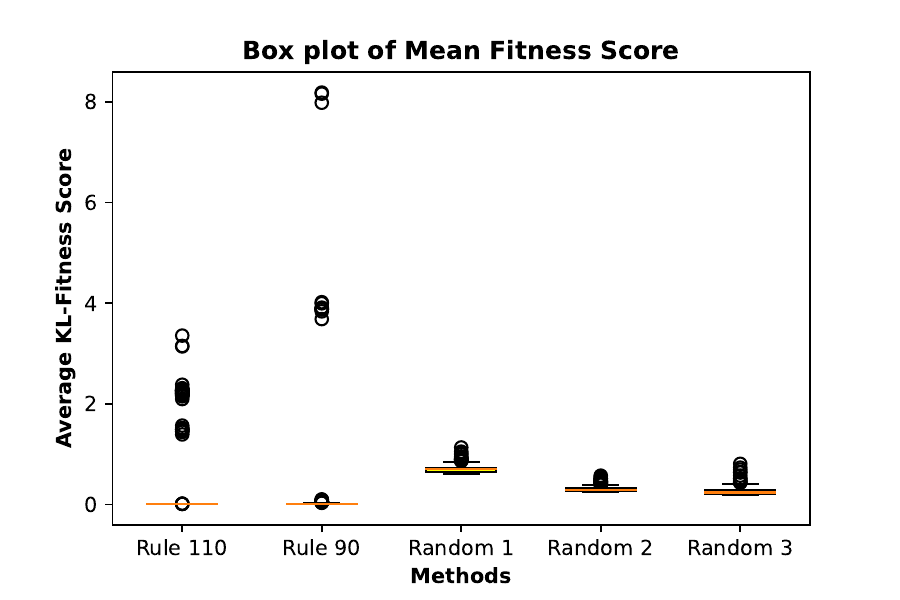}%
    \put(-150,0){\textbf{b)}} 
    \caption{\protect 
       \textbf{a)} Fitness scores for different sets of probabilities against the number of generations for the KL-fitness function.
       \textbf{b)} Best fitness scores per run for $D_{KL}$ fitness function for different sets of probabilities.
       The fitness scores of each gate are the average fitness scores of 50 runs.}
    \label{box_plot_exp3}
\end{figure}
\begin{figure}[t]
    \centering
    \includegraphics[width=0.7\textwidth]{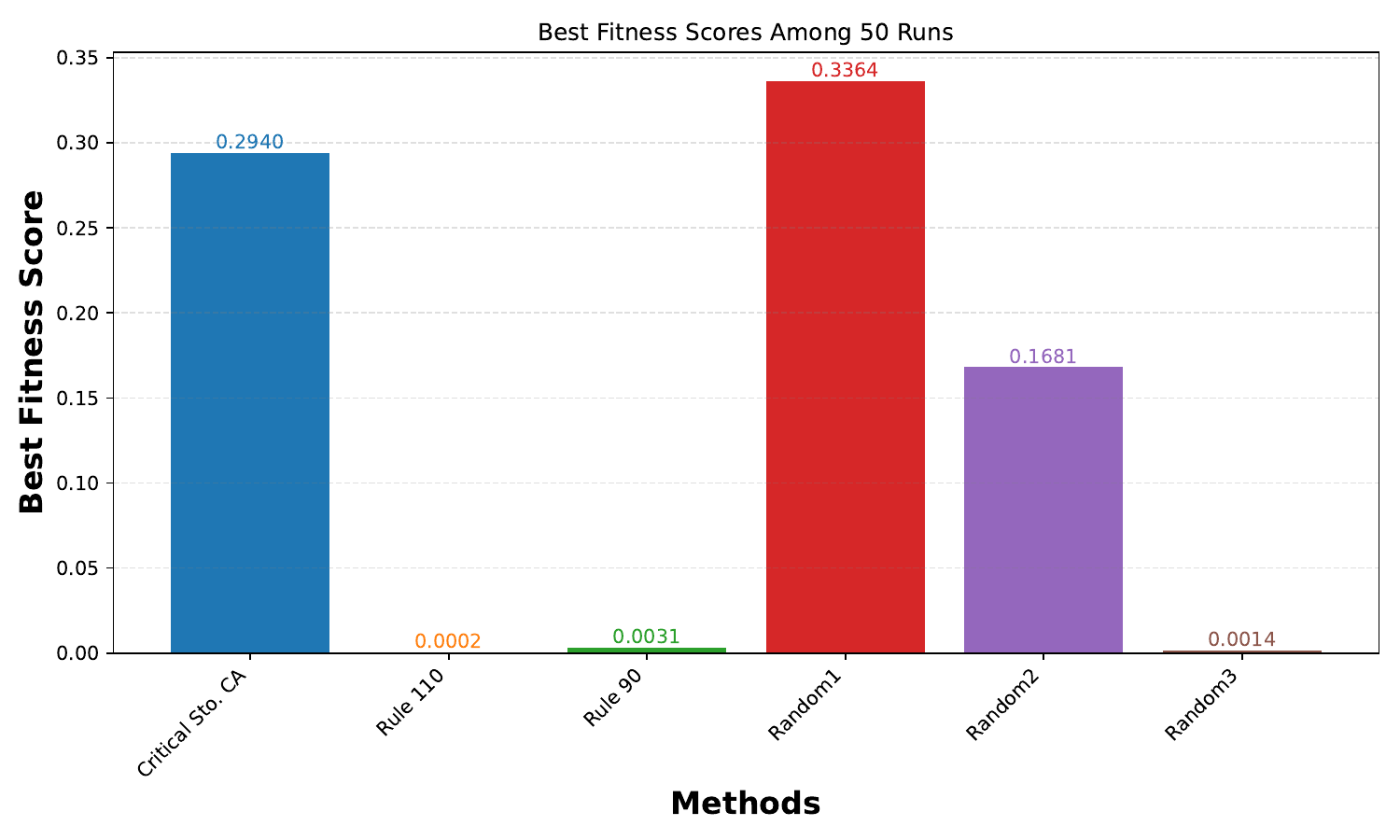}%
    \put(-260,5){\textbf{a)}}
    \\ 
    \includegraphics[width=1.0\textwidth]{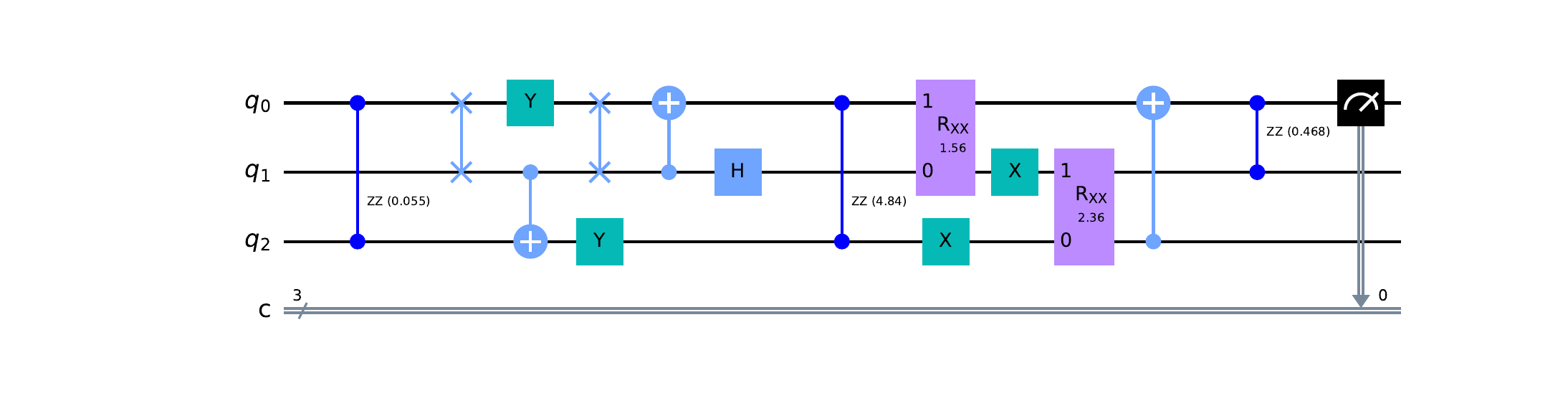}%
    \put(-300,5){\textbf{b)}} 
    \caption{\protect 
    \textbf{a)} Comparison of the best fitness scores across all 50 runs over 500 generations for different CA rules. 
    \textbf{b)} Visualization of the optimal quantum circuit: Achieving a fitness score of 0.294 using the $D_{KL}$ fitness function for Critical Stochastic Cellular Automaton. The circuit, composed of 15 gates with a 10\% mutation probability, represents the most exceptional outcome within the study, underscoring the success of our approach in circuit optimization.}
    \label{comparisionCA}
\end{figure}

The authors in~\cite{Pontes-Filho2020} have evolved a stochastic cellular automata model to reach criticality, a property of dynamical systems that allows them to do robust computations.
For each triad pattern, a probability has been calculated through a genetic algorithm for the central cell to have state 1. These probabilities are shown in Table~\ref{tab:probabilities_exp3}, second column. 

We start by considering the number of gates used, evolving quantum CAs with 3, 5, 10, 15, and 20 gates and 50 runs. The mutation probability is fixed at $10$ percent. The results are presented in Figure~\ref{box_plot}. They show that the best fitness score per run improves with increasing the number of gates up to 15 gates, after which a gradual increase is observed until the limit of 20 gates. Therefore, for experiments 2 and 3, $15$ gates are used. Furthermore, the results of experiments indicate that, while fitness scores initially improve with an increasing number of gates, the rate of progress diminishes as the number of gates continues to increase.

Next, we explore how the fitness changes with the value of the mutation probability. The goal here is to test the impact of the mutation on the fitness function of the different generations. We fixed the number of gates to 15 and used different mutation probabilities,  5\%, 10\%, 20\%, 30\%, and 50\%.

Figure~\ref{img:mutation_rate} illustrates the impact of different mutation rates on the $D_{KL}$ fitness score. Figure~\ref{img:mutation_rate} \textbf{a)} shows that all mutation rates result in a decrease in fitness scores over generations, indicating improvement. However, mutation rates of 5\%, 10\%, and 20\% stabilize more consistently compared to higher rates. Figure~\ref{img:mutation_rate} \textbf{b)} shows that while higher mutation rates (30\%, 50\%) lead to decreased fitness scores, they introduce greater variability and less stability. These results highlight that while higher mutation rates can still improve circuit efficiency, the optimal balance between exploration and stability is achieved at a 20\% mutation rate.

We now test the above approach with generic deterministic and stochastic cellular CAs. We fixed parameters by utilizing 15 gates and a 10\% mutation probability. Specifically, we examine the application of our methodology to Rule 90 and Rule 110 (deterministic rules with fixed probabilities of 0 or 1 for the eight neighborhoods), as well as to eight randomly generated probabilities, each repeated three times, serving as target probabilities. The corresponding target probabilities for each experimental condition are detailed in   Table~\ref{tab:probabilities_exp3}, while the ensuing outcomes are visually represented in Figure~\ref{box_plot_exp3}. This comprehensive analysis showcases the versatility and effectiveness of our approach across a spectrum of rule types and probabilistic settings.

The fitness scores obtained for the deterministic CAs, Rule 90 and Rule 110, are remarkably good, as shown in Figure~\ref{box_plot_exp3} \textbf{a)}. This highlights the efficiency of our approach to the synthesis of quantum circuits for deterministic CAs. Notably, the outcomes stemming from the stochastic CA approach, characterized by the incorporation of randomly generated probabilities over three separate iterations, exhibit some promise. The optimal fitness scores attained for Random1, Random2, and Random3 are 0.3364, 0.1681, and 0.0014, respectively. It is worth mentioning that while Random1 achieves a commendable fitness score of 0.3364, indicating its promising performance, there is potential for further refinement of the evolutionary process to reduce any existing disparities. A comprehensive evaluation of the fitness scores across different probability sets, as presented in Figure~\ref{box_plot_exp3} \textbf{b)}, to the number of generations, provides valuable insights into the performance trends for the deterministic CA models (Rule 90 and Rule 110), as well as the stochastic CA model utilizing randomly generated probabilities over three iterations (Random1, Random2, and Random3), all conducted with an initial set of 50 run conditions.

Figure~\ref{comparisionCA} \textbf{a)} illustrates the fitness scores corresponding to the diverse CA rules. The best fitness score of 0.294 was achieved by the critical CA. Notably, for two deterministic CA rules, namely Rule 90 and Rule 110, the fitness scores approximate zero, reflecting a deliberate pursuit of minimizing the absolute disparity between initial and final states. This trend aligns logically with the deterministic nature of these CA rules, where probability values exclusively assume binary states of 0 or 1. In contrast, the scenario changes when dealing with randomly generated stochastic CA rules. In this context, the fitness scores demonstrate variability across three distinct randomly assigned probabilities. Intriguingly, among these variations, Random 3 emerges as the front-runner, boasting the most favorable fitness scores. Remarkably, all three sets of randomly generated probabilities exhibit fitness values superior to that of the critical stochastic CA value. Nevertheless, to unravel the underlying mechanisms driving these observed phenomena, a comprehensive exploration beckons. This entails conducting a more extensive array of experiments, encompassing increased generations and a higher replication count. The intricate interplay of factors contributing to fitness score disparities between distinct randomly generated probabilities necessitates a more exhaustive investigation.
 
In Figure~\ref{comparisionCA} \textbf{b)}, we show the circuit 
that significantly outperforms the average outcomes of runs associated with the critical stochastic cellular automaton. Impressively, this particular realization achieved a fitness score of 0.294, as evaluated by the $D_{KL}$ fitness function outlined in Equation~(\ref{eq:FF2}). 

\subsection{Designing entangling quantum circuits}
\label{sec:entangle}

In this section, we employ an evolutionary algorithm to design quantum circuits that generate entangled states. To establish a proof of concept, we will specifically address systems of three, four, and five qubits, using the MW entanglement measure as a fitness function. We analyze these qubit numbers separately to explore how entanglement generation varies with system size, a critical factor in quantum circuit design. 
In the end, we discuss how the feasibility of our framework scales with the complexity of larger quantum circuits. Our experiments, conducted over 500 generations with 50 runs each, reveal how these factors influence entanglement optimization across different system sizes. 
\begin{figure}[t]
    \centering
    \includegraphics[width=.5\textwidth]{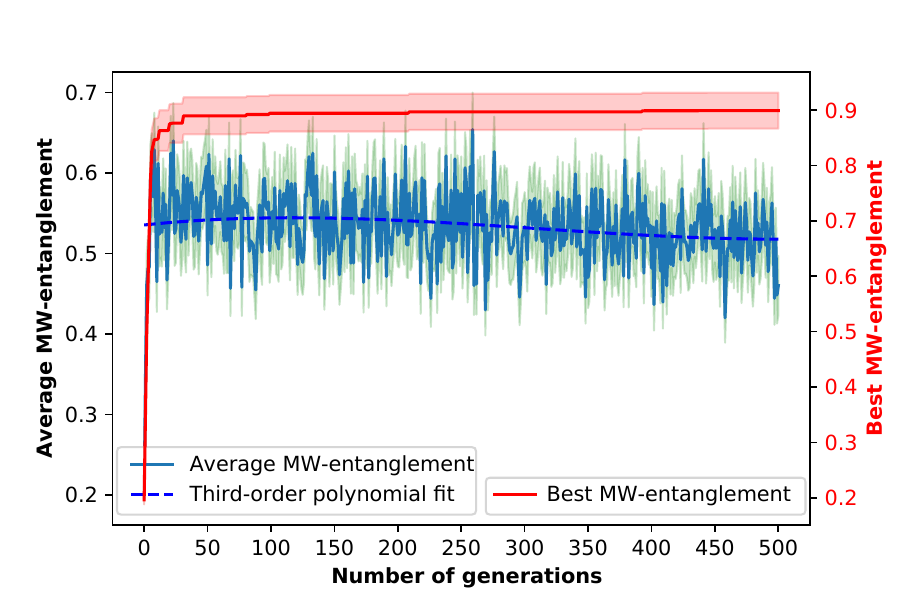}%
    \put(-150,0){\textbf{a)}} 
    \hfill
    \includegraphics[width=.5\textwidth]{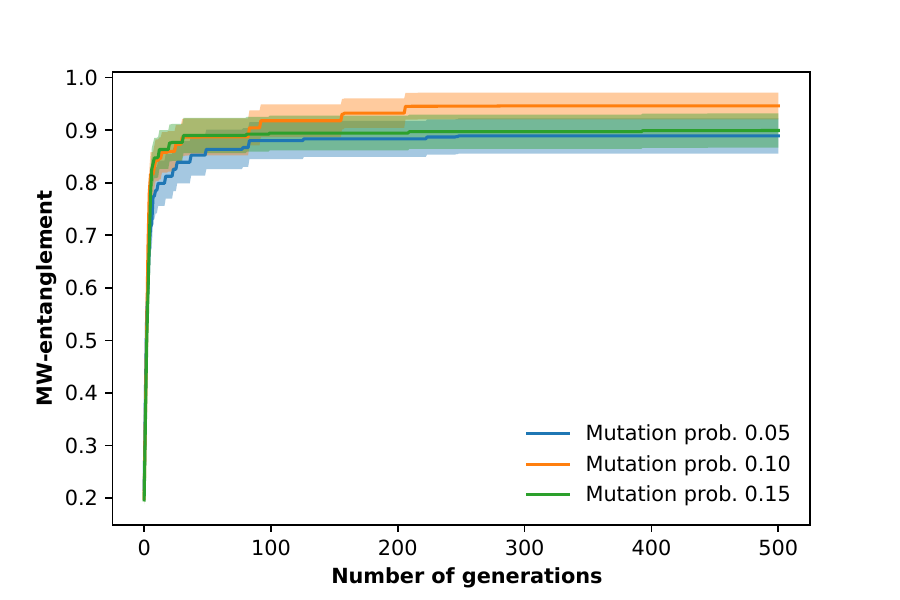}%
    \put(-150,0){\textbf{b)}} 
    \\%
    \includegraphics[width=.5\textwidth]{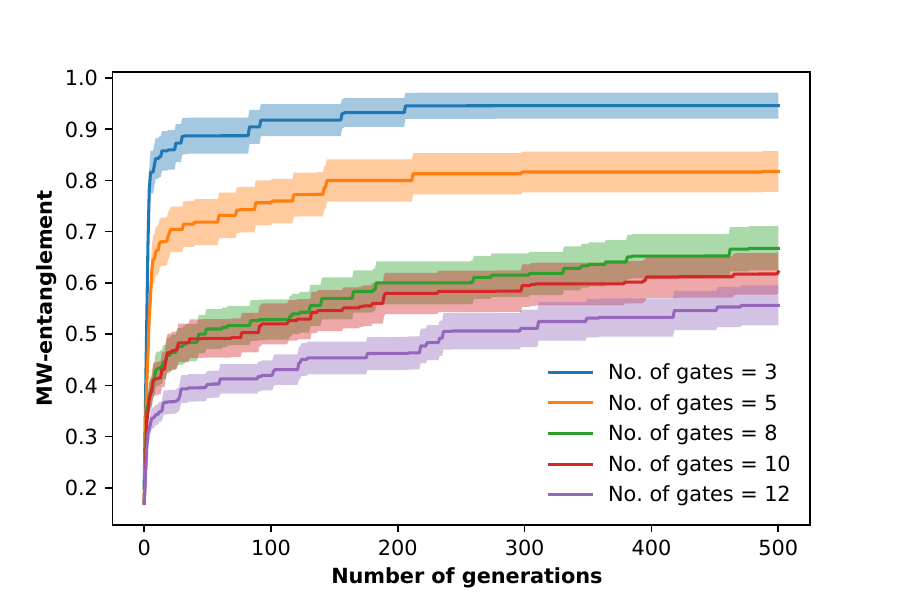}%
    \put(-150,0){\textbf{c)}} 
    \hfill
    \includegraphics[width=.5\textwidth]{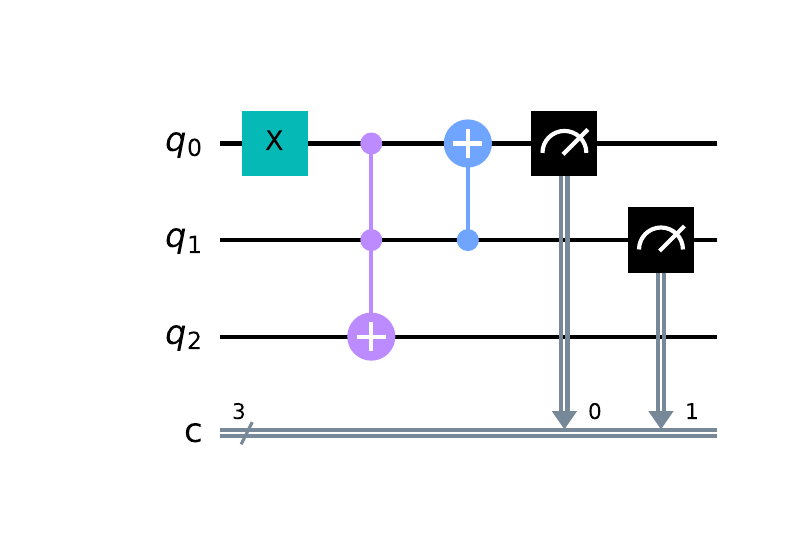}%
    \put(-150,0){\textbf{d)}} 
    \caption{\protect
    \textbf{a)} Evolutionary optimization of three-qubit quantum circuits with three gates using the MW entanglement measure~\cite{brennen2003observable} as the fitness function. 
    \textbf{b)} Plots show mean and best fitness across generations for different mutation rates: 5\%, including a third-order polynomial fit to the mean fitness. 
    \textbf{c)} Best fitness comparison with varying mutation rates and gate numbers. 
    \textbf{d)} Fitness outcomes for different gate numbers at a constant 10\% mutation probability, averaged over 50 runs with standard error bars. 
    An example circuit achieves a high fitness score of 0.999 with only three gates and 10\% mutation probability.}
    \label{fig:3qubitfigure}
\end{figure}


\subsubsection{Three-qubit circuits}
We start with a three-qubit case. 

Figure \ref{fig:3qubitfigure} presents the evolutionary optimization of a three-qubit quantum circuit across 500 generations, highlighting both average and peak fitness scores. We use three different mutation rates, 5\%, 10\%, and 15\%, and limit the number of gates in the circuits to  3, 5, 8, 10, and 12. The best fitness scores obtained for all mutation rates approached unity, indicating that the resulting circuits are indeed highly entangling. 

Since the best fitness scores for the quantum circuits approached unity, EAs indeed serve as a toolbox of methods to design highly entangling three-qubit circuits. For a three-qubit system (three lists of integer gates for each qubit), a mutation probability of 1/9 (11.11\%) is needed to replace at least one individual population consisting of nine integer lists. Therefore, a 10\% mutation rate, which can at least replace one individual population in each generation, can be considered an optimal probability for the mutation rate. Experimentally, the optimal mutation rate was found to be 10\%, which struck a balance between exploring different solutions in the search space and exploiting the best solutions as shown in Figure \ref{fig:3qubitfigure} \textbf{b)}. In addition, a lower mutation rate is capable of striking a balance between exploration and exploitation, leading to a higher average fitness score across all runs. However, the mutation rate of 10\% is more effective in building and generating the circuit that performs best in each run due to the higher degree of exploration it allows for the entanglement.

We also varied the number of gates in the quantum circuit in the range from 3 to 12, keeping the optimal mutation rate at 10\%. The result is that the fitness score decreased as the number of gates in the circuit increased, as shown in Figure \ref{fig:3qubitfigure} \textbf{c)}. These results suggest that as the quantum circuit becomes more complex with an increasing number of gates, the entanglement capability of the circuit is reduced, indicating that the circuit is less efficient at performing the desired task. This suggests that there may be a trade-off between circuit complexity and performance. Therefore, choosing the optimal number of gates can provide the best balance between the complexity of the quantum circuit and the performance.

The best three-qubit entangling circuit with three gates is shown in Figure~ \ref{fig:3qubitfigure} \textbf{d)}. It corresponds to the fitness score of 0.999. Similarly, a three-qubit quantum circuit with 12 gates also achieved a fitness value of 0.999. In \textbf{Supplementary Note III}, the detailed analyses of the state vector and reduced density matrices from optimal circuits in a 500-generation circuit are presented. Table \ref{table:compact_state_probabilities} in the Supplementary Note shows these details for the state vector probability of the respective density matrices. The methodologies used for calculating the probabilities of quantum states and the reduced density matrices for the three-qubit circuits, along with an assessment of their entanglement using the MW entanglement measure, are given in \textbf{Supplementary Note III}.

\begin{figure}[t]
    \centering    
    \includegraphics[width=0.5\textwidth]{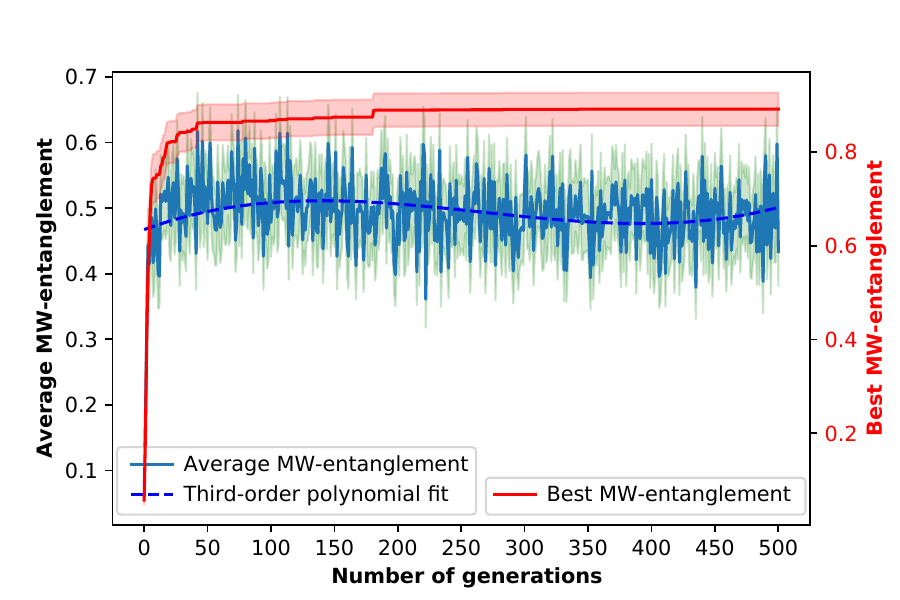}%
    \put(-150,0){\textbf{a)}} 
    \hfill
    \includegraphics[width=0.5\linewidth]{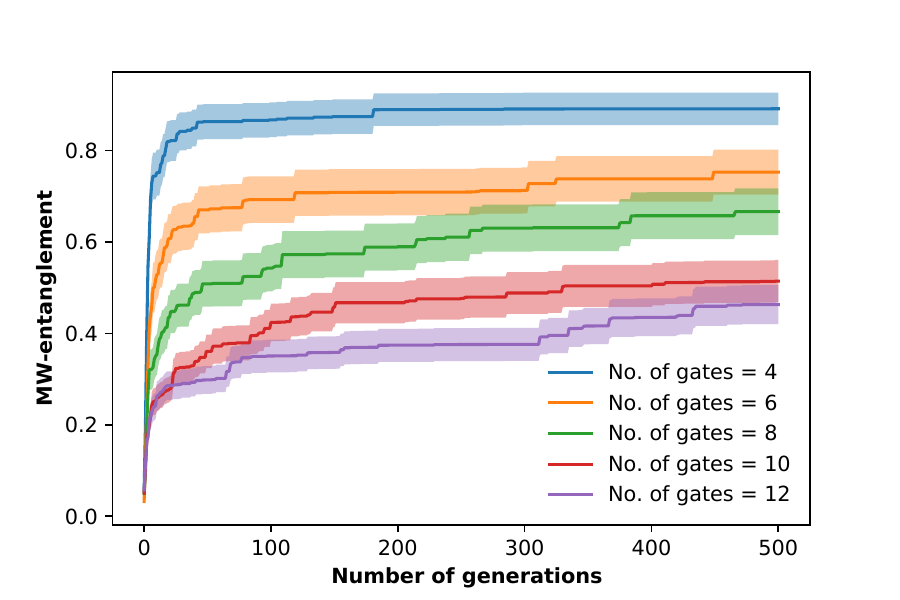}%
    \put(-150,0){\textbf{b)}} 
    \\ 
    \includegraphics[width=0.55\linewidth, height=0.32\textwidth]{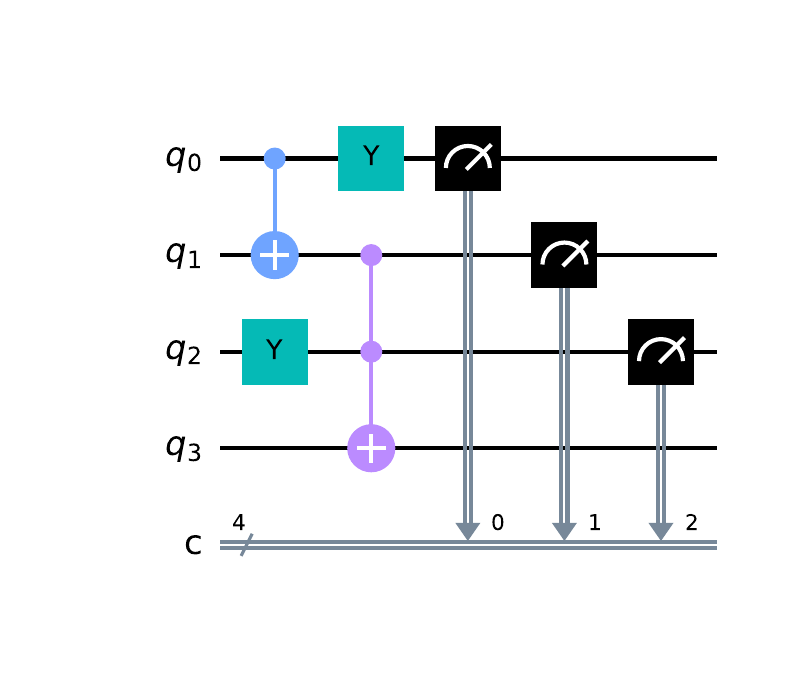}%
    \put(-150,15){\textbf{c)}} 

    \caption{\protect 
        \textbf{a)} Four-qubit quantum circuits with four gates and a 10\% mutation probability. The mean fitness (green line) and its shaded standard error, along with the mean of the best fitness (red line) and its shaded standard error, are plotted against the number of generations. 
        \textbf{b)} Comparison of the best fitness generated for five different numbers of gate sets in a four-qubit circuit using a 10\% mutation rate and 20 chromosomes. 
        \textbf{c)} Evolutionary generation of four-gate four-qubit circuits with MW-entanglement fitness scores using a 10\% mutation probability with a fitness score of 0.9996. }
    \label{fig:4Qubit_fitness}
\end{figure}

\subsubsection{Four and five-qubit circuits}

\begin{figure}[t]
    \centering
    \includegraphics[width=0.5\textwidth]{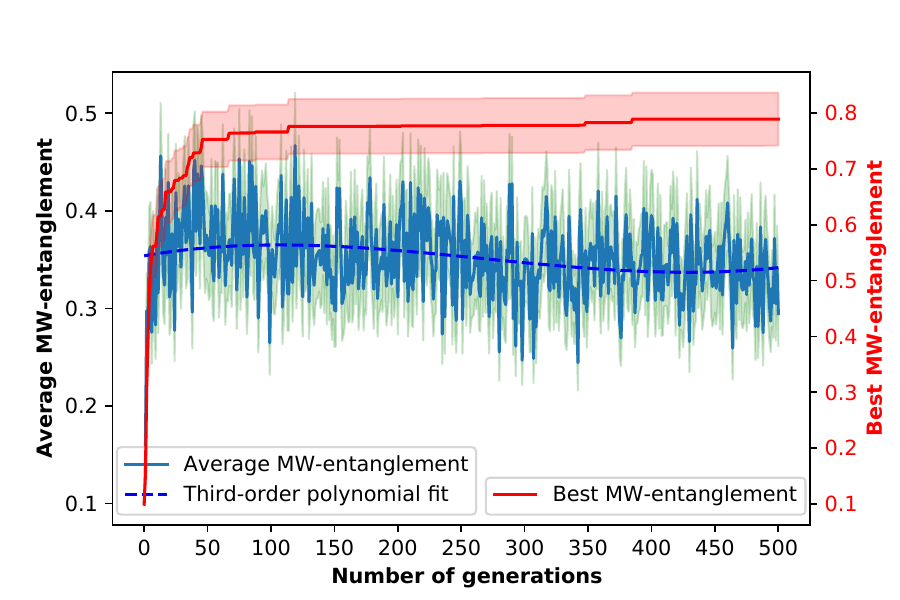}%
    \put(-150,0){\textbf{a)}} 
    \hfill
    \includegraphics[width=0.5\linewidth]{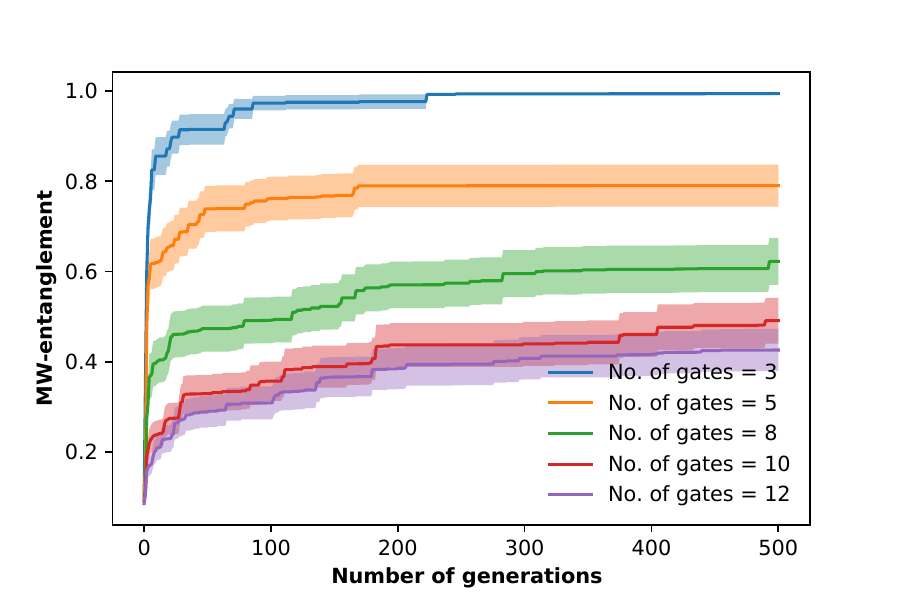}%
    \put(-150,0){\textbf{b)}} 
    \\ 
    \includegraphics[width=0.55\textwidth]{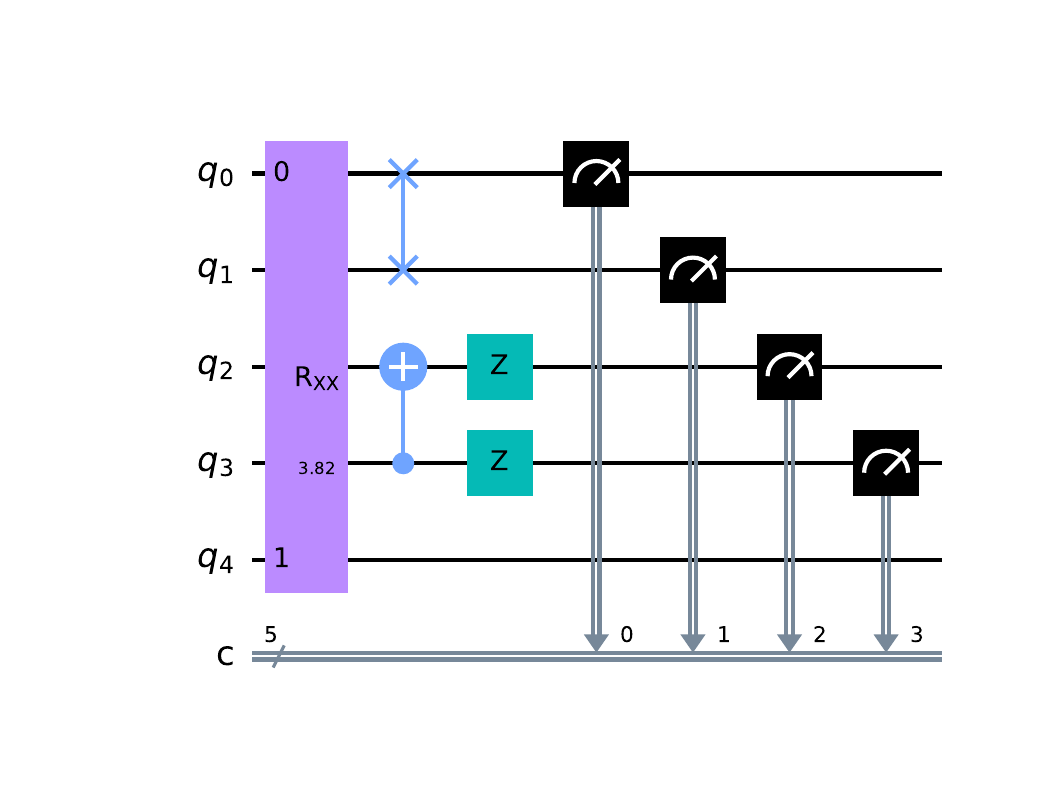}%
    \put(-150,15){\textbf{c)}} 
    \caption{\protect
        Evolutionary optimization of five-qubit quantum circuits: \textbf{a)} Optimization of five-qubit circuits with five gates and a 5\% mutation rate. The plots show the mean (green line) and best (red line) fitness scores, with standard error bars plotted against the number of generations. The blue dashed lines represent third-order polynomial fits to the mean fitness scores. \textbf{b)} Evolutionary optimization with varying gate numbers, all with a 5\% mutation rate. \textbf{c)} Fitness scores for circuits with five gates, achieving a score of 0.999.}

    \label{5Qent_gates}
\end{figure}

In this section, we extended the design of quantum circuits to advance our findings to include both four and five-qubit systems. The findings are shown in Figures \ref{fig:4Qubit_fitness} and \ref{5Qent_gates} for four and five-qubit quantum circuits, respectively. Notably, Figures~\ref{fig:4Qubit_fitness} \textbf{a)} and \ref{5Qent_gates} \textbf{a)} show the mean fitness (illustrated by a green line) alongside its standard error (denoted by a shaded area), in addition to the average of the best fitness (represented by a red line) and its standard error (also shaded). A third-order polynomial fit applied to the mean fitness is shown by a blue dashed line. These figure representations demonstrate that the average fitness score approximates 0.6, while the mean of the highest fitness score approaches 0.8.

Further experiment with an optimal mutation probability of 10\% and 5\% for four and five qubit circuits, respectively, was conducted, focusing on circuits with varying numbers of gates as shown in Figures~\ref{fig:4Qubit_fitness} \textbf{b)} and \ref{5Qent_gates} \textbf{b)}. Through an analysis of mutation rates of 3\%, 5\%, and 15\%, we found that increasing the mutation rate beyond 5\% results in a decrease in the average fitness value. This indicates that excessive randomness in the evolutionary process can have a negative impact on circuit performance. However, it was also observed that the average of the best fitness value was not affected by higher mutation rates. This highlights the importance of selecting an appropriate mutation rate for obtaining the quantum circuits of maximal entanglement.


The visualization of the quantum circuit with four and five qubits, incorporating four and five gates, respectively, achieving a fitness score of 0.999, is shown in Figures~\ref{fig:4Qubit_fitness} \textbf{c)} and \ref{5Qent_gates} \textbf{c)}. Additionally, the five-qubit quantum circuit with 12 gates yielded a fitness score of 0.199, signifying a decrement in entanglement capabilities with an increase in the number of gates. Detailed outcomes for the most entangled four and five-qubit quantum circuits, including the state vector and reduced density matrices, are elaborated in Section II of the Supplementary Note. These results pertain to the optimal five-qubit circuit identified across 500 generations, encompassing circuits with both five and twelve gates, as illustrated in Tables \ref{table:4qubit_combined_probabilities} and \ref{table:5qubit_combined_probabilities} \textbf{(Supplementary Note IV)}.

\subsubsection{The von Neumann entropy as a fitness function}

\begin{figure}[t]
    \centering
    \includegraphics[width=0.85\linewidth]{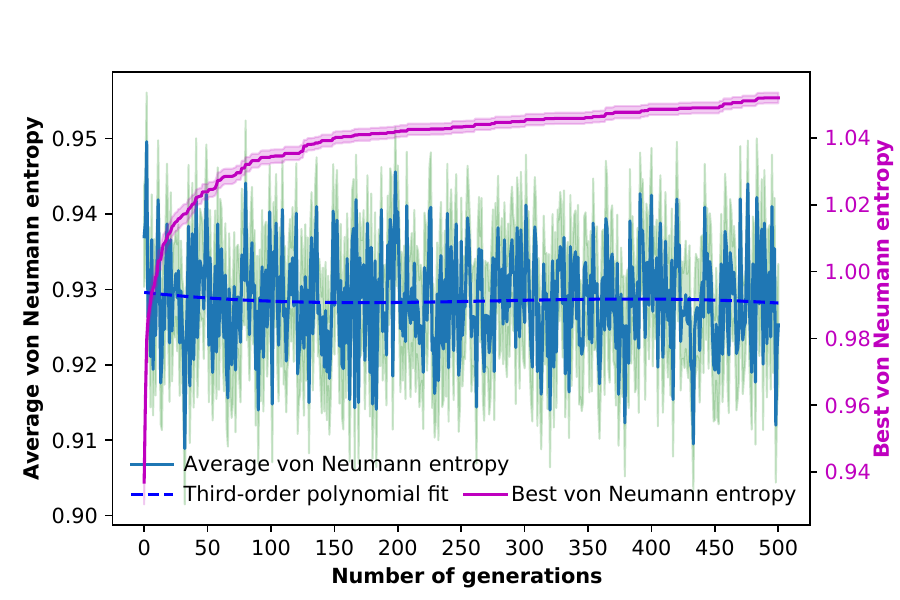}
    \caption{\protect Evolutionary optimization of three-qubit quantum circuits with 10\% mutation probability and ten gates, using von Neumann entanglement entropy as the fitness function. The plot shows the mean fitness (green line) and its shaded standard error, as well as the mean of the best fitness (magenta line) and its shaded standard error, against the number of generations. The blue dashed line represents the third-order polynomial fit to the mean fitness. The maximum fitness score obtained was 1.05557}.
    \label{vonn_neumann}
\end{figure}

In our experimental setup, we also incorporated von Neumann entropy within an evolutionary algorithm framework. The algorithm was configured for a three-qubit system equipped with ten gates, operating at a mutation rate of 10\%. This procedure was iteratively performed across 50 cycles, spanning 500 generations, with each generation consisting of 20 chromosomes. The results of this implementation, illustrating the effectiveness of von Neumann entropy in the optimization process, are graphically represented in Figure \ref{vonn_neumann}.

The theoretical maximum entropy for such a system is 2.999. Nevertheless, when implementing von Neumann entropy entanglement as a fitness function in the evolutionary algorithm, we obtained a maximum entropy value of 1.05557. This result has significant implications for the entanglement within the system. The quantification of entanglement can be achieved through the computation of von Neumann entropy for the reduced density matrix of any given subsystem. 

Our findings indicate that the implementation of von Neumann entropy in the three-qubit system successfully detected the presence of entanglement within the evolved three-qubit quantum circuit. This is consistent with the understanding that the entropy of entanglement is the von Neumann entropy of the reduced density matrix for any of the subsystems.
\section{Discussion and conclusions}
\label{sec:conclusions}
In this study, we explore the two examples of EAs in two distinct yet interconnected areas of quantum computing. Initially, we introduce a framework for generating quantum circuits that replicate specific cellular automata rules using genetic algorithms \cite{10.1007/978-3-031-14926-9_11}. The implementation code is available at the GitHub repository \url{https://github.com/Overskott/Quevo}. This framework effectively evolves quantum circuits for various CA rules, from deterministic to stochastic types. We improved our original framework by increasing the number of generations and runs, enhancing performance, and achieving more consistent fitness function outcomes. This initial investigation set the stage for more nuanced applications of EAs in quantum circuit design, suggesting future research paths. 

Building upon this groundwork, we have demonstrated the effectiveness of EAs in generating highly entangled quantum circuits, emphasizing their role in calculating reduced density matrices for three-, four-, and five-qubit systems. 
These circuits exhibited significant entanglement, as quantified by the MW entanglement measure. 
At the same time, as the number of qubits increases, so does the complexity of entanglement—from well-known three-qubit states like GHZ or W states \cite{UCHIDA20152698} to more intricate multipartite structures in four- and five-qubit systems \cite{MISHRA2024100230}. 
This progression allows us to assess the scalability of our evolutionary approach, identify optimal circuit configurations, and evaluate the trade-off between circuit complexity (e.g., gate count) and entanglement performance. Moreover, they are of practical relevance, since studying three to five qubits aligns with the capabilities of near-term quantum hardware, where noise and decoherence limit the practical utility of large qubit counts.

However, while the sizes chosen in our work are sufficient for proof-of-concept demonstrations, our findings can now be extended toward complexity levels beyond the sizes addressed in this work. 
Although with limited access, current quantum hardware, such as IBM’s devices with several hundred qubits \cite{abughanem2024ibmquantumcomputersevolution}, highlights the need to assess how our methods perform beyond this range. Scaling to larger systems, with 10–20 qubits or more, exponentially increases the search space for evolutionary optimization,
which is already slow and computationally costly by design.
This scalability challenge, when combining quantum and EAs, is compounded by the trade-off between circuit complexity and entanglement performance. In this respect, our results show that entanglement capability decreases with increasing circuit depth in small systems. Specifically, a five-qubit circuit with 12 gates scores 0.199 versus 0.999 with five gates. In larger systems, this trade-off is likely to intensify, as deeper circuits may be required to achieve entanglement across more qubits, yet excessive depth could further degrade performance due to accumulated errors and decoherence. 

To stress that, our focus on small system sizes is motivated by the constraints of near-term quantum hardware, where noise and decoherence limit the practical use of large qubit counts, rendering optimized small circuits highly relevant for current applications. Nevertheless, we acknowledge that this restricts the immediate applicability of our findings to larger-scale quantum computing scenarios. To address this, future work could leverage hybrid classical-quantum optimization techniques or execute the EAs directly on quantum hardware, potentially offering speed advantages and enabling scalability to larger systems. Such advancements could extend our approach to larger, more realistic quantum systems, enhancing their practical applicability in settings where hundreds of qubits are available.

Another key finding is the identification of an optimal mutation rate for quantum circuits across different qubit sizes and gate counts, marking a significant step in optimizing entanglement generation. To note that the $D_{KL}$ fitness score measures the discrepancy between the distributions of the output states and the target states. In our case, these distributions are discrete, each corresponding to the eight different initial states of the three qubits. If the distributions completely match the fitness function, the $D_{KL}$ value is zero. A decreasing fitness value close to zero indicates an increase in the efficiency/entanglement of the circuits.
Our findings show that with mutation rates up to 20\%, the optimization process effectively explores the solution space, leading to better circuit designs with lower fitness scores. However, mutation rates beyond 20\% introduce excessive randomness, which disrupts the optimization process and results in higher fitness scores, indicating less efficient/entangled circuits. 
All mutation rates result in a decrease in fitness scores over generations, indicating improvement. However, mutation rates of 5\%, 10\%, and 20\% stabilize more consistently compared to higher rates. The right panel box plots show that while higher mutation rates (30\%, 50\%) lead to decreased fitness scores, they introduce greater variability and less stability. These results highlight that while higher mutation rates can still improve circuit efficiency, the optimal balance between exploration and stability during the "evolution" of the algorithm is achieved at a 20\% mutation rate.

Finally, some last points for discussion. This paper opens avenues for future investigations into advanced EAs and alternative fitness functions, such as Von Neumann entropy \cite{Mendes_Santos_2020} and Schmidt measures \cite{Eisert_2001,https://doi.org/10.48550/arxiv.quant-ph/0505149}, to maximize entanglement in more complex quantum circuits.  Future work could explore the scalability of these methods to larger quantum circuits, composed of more than five qubits.  The fitness function here introduced, measuring the entanglement between qubits, could be extended to a multi-dimensional optimization scheme, incorporating also measures of error and noise in quantum circuits.  Moreover, it would also be of interest to adapt the method proposed here and integrate it into hybrid optimization approaches.
Future research could explore the maximization of entanglement in quantum circuits, drawing inspiration from recent advancements in benchmarking highly entangled states on a 60-atom analog quantum simulator\cite{Storz2023} and leveraging non-locality as a resource in quantum computing demonstrated by loophole-free Bell inequality violations with superconducting circuits\cite{Shaw2024}. Ultimately, applying our optimized quantum circuits to real-world problems, such as quantum cryptography\cite{Nadlinger2022,PQCrypto2024}, quantum simulation\cite{PQCrypto2024,Shaw2024}, and quantum machine learning\cite{Schatzki2021EntangledDF}—could further demonstrate their practical utility and effectiveness. In gate-model quantum computing frameworks, such as scalable distributed systems\cite{Gyongyosi2021} or circuit depth reduction efforts\cite{Gyongyosi2020}, our EAs could enhance efficiency by minimizing gate counts while maximizing entanglement, aligning with near-term hardware needs. Similarly, in a quantum internet context\cite{10.1145/3524455}, our highly entangled circuits could support protocols like quantum key distribution or teleportation, contributing to distributed quantum networks. Also, one could explore the scalability of our approach to larger quantum circuits, building on the experimental advancements demonstrated by Chen et al. \cite{chiu2025integration} and the theoretical frameworks proposed by Arrighi et al. \cite{arrighi2022quantum}. Additionally, integrating noise resilience techniques into our evolutionary algorithm framework could further enhance the stability and performance of quantum circuits in noisy environments.




\section*{Data and code availability}
The datasets and the implementation code used and/or analyzed during the current study are available in the QUEVO1 repository, \url{https://github.com/shailendrabhandari/QUEVO1}. 

\section*{Acknowledgements}

We extend our gratitude to Sebastian T. Overskott for making the \href{https://github.com/Overskott/Quevo}{Quevo} framework available, which we have adapted for the entanglement assessment in our study. 

\section*{Author contributions statement}

S.B. conducted the experiments and performed the simulations. All authors have analyzed the results and written the manuscript. S.N., S.D., and P.G.L. have revised the text, and P.G.L. coordinated the overall project.
%
\bibliography{references}

\begin{thebibliography}{10}

\bibitem{abughanem2024ibmquantumcomputersevolution}
M.~AbuGhanem, (2024).
\newblock Ibm quantum computers: Evolution, performance, and future directions.

\bibitem{ACAMPORA2021542}
Giovanni Acampora and Autilia Vitiello.
\newblock (2021).
\newblock {Implementing evolutionary optimization on actual quantum processors}.
\newblock {\em Information Sciences}, 575:542--562.

\bibitem{Addala2025}
V.~L. Addala, S.~Ge, and S.~Krastanov.
\newblock (2025).
\newblock Faster-than-clifford simulations of entanglement purification circuits and their full-stack optimization.
\newblock {\em npj Quantum Information}, 11:12.

\bibitem{Amal2022}
R.~S. Amal and J.~Solomon Ivan.
\newblock (2022).
\newblock {A quantum genetic algorithm for optimization problems on the Bloch sphere}.
\newblock {\em Quantum Information Processing}, 21(2):1--29.

\bibitem{arrighi2022quantum}
Pablo Arrighi, Amélia Durbec, and Matt Wilson, (2022).
\newblock Quantum networks theory.

\bibitem{arriola2020bio}
Victor~Manuel Arriola-Rios, Carlos~A Cruz-Ramos, and Gustavo Arroyo-Figueroa.
\newblock (2020).
\newblock Bio-inspired optimization for quantum computing: a review.
\newblock {\em arXiv preprint arXiv:2010.06437}.

\bibitem{ballinas2022hybrid}
Edgar Ballinas and Osvaldo Montiel.
\newblock (2022).
\newblock Hybrid quantum genetic algorithm with adaptive rotation angle for the 0-1 knapsack problem in the ibm qiskit simulator.
\newblock {\em Soft Computing}.

\bibitem{Ballinas2022}
Eduardo Ballinas and Oscar Montiel.
\newblock (2022).
\newblock Hybrid quantum genetic algorithm with adaptive rotation angle for the 0-1 knapsack problem in the ibm qiskit simulator.
\newblock {\em Soft Computing}, 26(18):9451--9463.

\bibitem{10.1348/000711010X522227}
Dmitry~I. Belov and Ronald~D. Armstrong.
\newblock (2011).
\newblock Distributions of the kullback–leibler divergence with applications.
\newblock {\em British Journal of Mathematical and Statistical Psychology}, 64(2):291--309.

\bibitem{benenti2007principles}
Giuliano Benenti, Giulio Casati, and Giuliano Strini.
\newblock (2007).
\newblock {\em Principles of Quantum Computation and Information}.
\newblock WORLD SCIENTIFIC.

\bibitem{PhysRevLett.70.1895}
C.~H. Bennett, G.~Brassard, C.~Crépeau, R.~Jozsa, A.~Peres, and W.~K. Wootters.
\newblock (1993).
\newblock Teleporting an unknown quantum state via dual classical and {{Einstein}}-{{Podolsky}}-{{Rosen}} channels.
\newblock {\em Physical Review Letters}, 70(13):1895--1899.

\bibitem{PhysRevA.54.3824}
Charles~H. Bennett, David~P. DiVincenzo, John~A. Smolin, and William~K. Wootters.
\newblock (11 1996).
\newblock Mixed-state entanglement and quantum error correction.
\newblock {\em Phys. Rev. A}, 54:3824--3851.

\bibitem{10.1007/978-3-031-14926-9_11}
Shailendra Bhandari, Sebastian Overskott, Ioannis Adamopoulos, Pedro~G. Lind, Sergiy Denysov, and Stefano Nichele.
\newblock (2022).
\newblock Evolving quantum circuits to implement stochastic and deterministic cellular automata rules.
\newblock In {\em International Conference on Cellular Automata for Research and Industry}, pages 119--129. Springer.

\bibitem{Biamonte2017}
Jacob Biamonte, Peter Wittek, Nicola Pancotti, Patrick Rebentrost, Nathan Wiebe, and Seth Lloyd.
\newblock (2017).
\newblock Quantum machine learning.
\newblock {\em Nature}, 549(7671):195--202.

\bibitem{brennen2003observable}
Gavin~K. Brennen, (2003).
\newblock An observable measure of entanglement for pure states of multi-qubit systems.

\bibitem{Brown_2023}
Jonathon Brown, Mauro Paternostro, and Alessandro Ferraro.
\newblock (01 2023).
\newblock Optimal quantum control via genetic algorithms for quantum state engineering in driven-resonator mediated networks.
\newblock {\em Quantum Science and Technology}, 8(2):025004.

\bibitem{PhysRevA.100.022342}
Alba Cervera-Lierta, Jos\'e~Ignacio Latorre, and Dardo Goyeneche.
\newblock (08 2019).
\newblock Quantum circuits for maximally entangled states.
\newblock {\em Phys. Rev. A}, 100:022342.

\bibitem{chiu2025integration}
Kuei-Lin Chiu, Youyi Chang, Avishma~J. Lasrado, Cheng-Han Lo, Yung-Hsiang Chen, Tao-Yi Hsu, Yen-Chih Chen, Yi-Chen Tsai, Samina, Yen-Hsiang Lin, and Chung-Ting Ke.
\newblock (2025).
\newblock Integration of graphene-based superconducting quantum circuits in 3d cavity.
\newblock {\em Physical Review Applied}, 23:034059.

\bibitem{EisertJ2010Alft}
J.~Eisert, M.~Cramer, and M.~B. Plenio.
\newblock (February 2010).
\newblock Colloquium: Area laws for the entanglement entropy.
\newblock {\em Reviews of Modern Physics}, 82(1):277–306.

\bibitem{https://doi.org/10.48550/arxiv.quant-ph/0505149}
J.~Eisert and D.~Gross, (2005).
\newblock Multi-particle entanglement.

\bibitem{Eisert_2001}
Jens Eisert and Hans~J. Briegel.
\newblock (07 2001).
\newblock Schmidt measure as a tool for quantifying multiparticle entanglement.
\newblock {\em Physical Review A}, 64(2).

\bibitem{doi:10.1080/14789940801912366}
V.~Murg F.~Verstraete and J.I. Cirac.
\newblock (2008).
\newblock Matrix product states, projected entangled pair states, and variational renormalization group methods for quantum spin systems.
\newblock {\em Advances in Physics}, 57(2):143--224.

\bibitem{friis2019entanglement}
Nicolai Friis, Giuseppe Vitagliano, Mehul Malik, Norbert L{"u}tkenhaus, Anders~S S{\o}rensen, Michalis Skotiniotis, and Animesh Datta.
\newblock (2019).
\newblock Entanglement certification from theory to experiment.
\newblock {\em Nat Rev Phys}, 1(2):72--87.

\bibitem{RevModPhys.74.145}
Nicolas Gisin, Gr\'egoire Ribordy, Wolfgang Tittel, and Hugo Zbinden.
\newblock (Mar 2002).
\newblock Quantum cryptography.
\newblock {\em Rev. Mod. Phys.}, 74:145--195.

\bibitem{Gyongyosi2020}
Laszlo Gyongyosi and Sandor Imre.
\newblock (2020).
\newblock Circuit depth reduction for gate-model quantum computers.
\newblock {\em Scientific Reports}, 10:11229.
\newblock Received 19 September 2019, Accepted 13 May 2020, Published 08 July 2020.

\bibitem{Gyongyosi2021}
Laszlo Gyongyosi and Sandor Imre.
\newblock (2021).
\newblock Scalable distributed gate-model quantum computers.
\newblock {\em Scientific Reports}, 11:5172.
\newblock Received 18 May 2020, Accepted 30 October 2020, Published 26 February 2021.

\bibitem{10.1145/3524455}
Laszlo Gyongyosi and Sandor Imre.
\newblock (July 2022).
\newblock Advances in the quantum internet.
\newblock {\em Commun. ACM}, 65(8):52–63.

\bibitem{articleQuantumEnranglement}
Kuk-Hyun Han and Jong-Hwan Kim.
\newblock (01 2003).
\newblock Quantum-inspired evolutionary algorithm for a class of combinatorial optimization.
\newblock {\em Evolutionary Computation, IEEE Transactions on}, 6:580 -- 593.

\bibitem{PhysRevLett.92.187901}
Aram Harrow, Patrick Hayden, and Debbie Leung.
\newblock (May 2004).
\newblock Superdense coding of quantum states.
\newblock {\em Phys. Rev. Lett.}, 92:187901.

\bibitem{PhysRevA.71.032313}
Karol \ifmmode~\dot{Z}\else \.{Z}\fi{}yczkowski and Hans-J\"urgen Sommers.
\newblock (03 2005).
\newblock Average fidelity between random quantum states.
\newblock {\em Phys. Rev. A}, 71:032313.

\bibitem{jones2022small}
E.~B. Jones, L.~E. Hillberry, M.~T. Jones, {\it et~al.}
\newblock (2022).
\newblock Small-world complex network generation on a digital quantum processor.
\newblock {\em Nature Communications}, 13:4483.

\bibitem{Longo_2020}
Roberto Longo and Feng Xu.
\newblock (2 2020).
\newblock Von neumann entropy in {QFT}.
\newblock {\em Communications in Mathematical Physics}, 381(3):1031--1054.

\bibitem{Lucas2019}
Simon~M. Lucas and Vanessa Volz.
\newblock (07 2019).
\newblock Tile pattern kl-divergence for analysing and evolving game levels.
\newblock {\em Proceedings of the Genetic and Evolutionary Computation Conference}.

\bibitem{Martin2015}
Fernando Martín, Luis Moreno, Santiago Garrido, and Dolores Blanco.
\newblock (2015).
\newblock Kullback-leibler divergence-based differential evolution markov chain filter for global localization of mobile robots.
\newblock {\em Sensors}, 15(9):23431--23458.

\bibitem{Mendes_Santos_2020}
T~Mendes-Santos, G~Giudici, R~Fazio, and M~Dalmonte.
\newblock (01 2020).
\newblock Measuring von neumann entanglement entropies without wave functions.
\newblock {\em New Journal of Physics}, 22(1):013044.

\bibitem{doi:10.1063/1.1497700}
David~A Meyer and Nolan~R Wallach.
\newblock (2002).
\newblock {Global entanglement in multiparticle systems}.
\newblock {\em Journal of Mathematical Physics}, 43(9):4273--4278.

\bibitem{MISHRA2024100230}
Ansh Mishra, Soumik Mahanti, Abhinash~Kumar Roy, and Prasanta~K. Panigrahi.
\newblock (2024).
\newblock Geometric genuine multipartite entanglement for four-qubit systems.
\newblock {\em Physics Open}, 20:100230.

\bibitem{Mooney2019}
Gary~J. Mooney, Charles~D. Hill, and Lloyd~C.L. Hollenberg.
\newblock (2019).
\newblock {Entanglement in a 20-Qubit Superconducting Quantum Computer}.
\newblock {\em Scientific Reports}, 9(1):1--8.

\bibitem{Nadlinger2022}
David~P. Nadlinger, Peter Drmota, Ben~C. Nichol, {\it et~al.}
\newblock (2022).
\newblock Experimental quantum key distribution certified by bell's theorem.
\newblock {\em Nature}, 607:682--686.

\bibitem{nielsen_quantum_2002}
Michael~A. Nielsen and Isaac~L. Chuang.
\newblock (2002).
\newblock {\em Quantum Computation and Quantum Information}.
\newblock Cambridge University Press, Cambridge.

\bibitem{nielsen_chuang_2010}
Michael~A. Nielsen and Isaac~L. Chuang.
\newblock (2010).
\newblock {\em Quantum Computation and Quantum Information: 10th Anniversary Edition}.
\newblock Cambridge University Press.

\bibitem{Pontes-Filho2020}
Sidney Pontes-Filho, Pedro Lind, Anis Yazidi, Jianhua Zhang, Hugo Hammer, Gustavo~B.M. Mello, Ioanna Sandvig, Gunnar Tufte, and Stefano Nichele.
\newblock (2020).
\newblock {A neuro-inspired general framework for the evolution of stochastic dynamical systems: Cellular automata, random Boolean networks and echo state networks towards criticality}.
\newblock {\em Cognitive Neurodynamics}, 14(5):657--674.

\bibitem{PQCrypto2024}
Markku-Juhani Saarinen and Daniel Smith-Tone, editors.
\newblock (2024).
\newblock {\em Post-Quantum Cryptography}, volume 14771 of {\em Lecture Notes in Computer Science}.
\newblock Springer Cham, 1 edition.

\bibitem{Schatzki2021EntangledDF}
Louis Schatzki, Andrew Arrasmith, Patrick~J. Coles, and Mar{\'i}a Cerezo.
\newblock (2021).
\newblock Entangled datasets for quantum machine learning.
\newblock {\em ArXiv}, abs/2109.03400.

\bibitem{2014}
Maria Schuld, Ilya Sinayskiy, and Francesco Petruccione.
\newblock (10 2014).
\newblock An introduction to quantum machine learning.
\newblock {\em Contemporary Physics}, 56(2):172–185.

\bibitem{Shaw2024}
A.L. Shaw, Z.~Chen, J.~Choi, {\it et~al.}
\newblock (2024).
\newblock Benchmarking highly entangled states on a 60-atom analogue quantum simulator.
\newblock {\em Nature}, 628:71--77.

\bibitem{Protocols}
Chitra Shukla, Anindita Banerjee, and Anirban Pathak.
\newblock (03 2014).
\newblock Protocols and quantum circuits for implementing entanglement concentration in cat state, ghz-like state and 9 families of 4-qubit entangled states.
\newblock {\em Quantum Information Processing}, 14.

\bibitem{Sim_20191}
Sukin Sim, Peter~D. Johnson, and Al{\'{a}}n Aspuru-Guzik.
\newblock (oct 2019).
\newblock Expressibility and entangling capability of parameterized quantum circuits for hybrid quantum classical algorithms.
\newblock {\em Advanced Quantum Technologies}, 2(12):1900070.

\bibitem{Storz2023}
S.~Storz, J.~Schär, A.~Kulikov, {\it et~al.}
\newblock (2023).
\newblock Loophole-free bell inequality violation with superconducting circuits.
\newblock {\em Nature}, 617:265--270.

\bibitem{Dancing}
R.S. Sutor.
\newblock (2019).
\newblock {\em Dancing with Qubits: How Quantum Computing Works and how it Can Change the World}.
\newblock Expert Insight. Packt Publishing.

\bibitem{https://doi.org/10.48550/arxiv.2302.01303}
Leo Sünkel, Darya Martyniuk, Denny Mattern, Johannes Jung, and Adrian Paschke, (2023).
\newblock Ga4qco: Genetic algorithm for quantum circuit optimization.

\bibitem{sunkel2023ga4qcogeneticalgorithmquantum}
Leo Sünkel, Darya Martyniuk, Denny Mattern, Johannes Jung, and Adrian Paschke, (2023).
\newblock Ga4qco: Genetic algorithm for quantum circuit optimization.

\bibitem{sunkel2023ga4qco}
Leo Sünkel, Darya Martyniuk, Denny Mattern, Johannes Jung, and Adrian Paschke, (2023).
\newblock Ga4qco: Genetic algorithm for quantum circuit optimization.

\bibitem{Tacchino2019}
Francesco Tacchino, Chiara Macchiavello, Dario Gerace, and Daniele Bajoni.
\newblock (2019).
\newblock {An artificial neuron implemented on an actual quantum processor}.
\newblock {\em npj Quantum Information}, 5(1):1--8.

\bibitem{UCHIDA20152698}
Gabriele Uchida, Reinhold~A. Bertlmann, and Beatrix~C. Hiesmayr.
\newblock (2015).
\newblock Entangled entanglement: A construction procedure.
\newblock {\em Physics Letters A}, 379(42):2698--2703.

\bibitem{2008}
F.~Verstraete, V.~Murg, and J.I. Cirac.
\newblock (03 2008).
\newblock Matrix product states, projected entangled pair states, and variational renormalization group methods for quantum spin systems.
\newblock {\em Advances in Physics}, 57(2):143–224.

\bibitem{PhysRevLett.91.147902}
Guifr\'e Vidal.
\newblock (10 2003).
\newblock Efficient classical simulation of slightly entangled quantum computations.
\newblock {\em Phys. Rev. Lett.}, 91:147902.

\bibitem{HWeinfurter_1994}
H.~Weinfurter.
\newblock (mar 1994).
\newblock Experimental bell-state analysis.
\newblock {\em Europhysics Letters}, 25(8):559.

\bibitem{WolframStephen1984Caam}
Stephen Wolfram.
\newblock (1984).
\newblock {Cellular automata as models of complexity}.
\newblock {\em Nature (London)}, 311(5985):419--424.

\bibitem{articleQE}
Bertrand Wong.
\newblock (12 2019).
\newblock On quantum entanglement.
\newblock {\em International Journal of Automatic Control System}, 5:1--7.

\end{thebibliography}

\appendix

\section*{Supplementary Note I: Implementation of an evolutionary quantum algorithm}
\label{supp1implementation}
Our approach was implemented in Python and resulted in a Python module called \href{https://github.com/Overskott/Quevo}{QUEVO}. This module consists of three classes: \verb!Chromosome!, \verb!Generation!, and \verb!Circuit!. The \verb!Chromosome! and the \verb!Generation! classes are part of the genetic algorithm, while the \verb!Circuit! class is responsible for generating and simulating quantum circuits.


The \verb!Chromosome! class is the core of the genetic algorithm used for generating quantum circuits. It handles the integer representation of the gates and their connections. The class contains a list of integers and functions to generate and mutate the list. It also handles the initialization of the population and the evolution of the population into a new one. Mutations in the \verb!Chromosome! class can occur in two different ways: replacing gates from the pool of gates in the chromosome with a randomly generated new one or replacing the chromosome to generate the four best parents. The class is also responsible for checking the chromosome for gates that connect multiple qubits. If the gate has an invalid connection, meaning that it is connected to itself through the randomly generated integers, the class generates a valid configuration randomly.

The \verb!Generation! class is another important class in the QUEVO module, responsible for managing the population of chromosomes. It is a collection of chromosomes that undergo evolution for a specified number of generations. After each evolution step, changes in the chromosomes in the generation occur by selecting a fixed number of chromosomes as elite and allowing the rest of them to evolve further. In each evolution step, the chromosomes are evaluated with the fitness function, and the fittest chromosomes become the parents for the next generation. The rate of the chromosomes is reset for the initial chromosomes. The \verb!Generation! class stores a generation of chromosomes, the fitness associated with each chromosome, methods for running and retrieving fitness for two different fitness functions, and functions for printing. It provides methods for performing selection, crossover, and mutation to generate a new population of chromosomes. 

The \verb!Circuit! class within the QUEVO module, which utilizes IBM's Qiskit (\url{https://qiskit.org/}), plays a crucial role in generating and managing quantum circuits. It is designed to create Qiskit quantum circuits from string representations, simulate these circuits, perform measurements, and visualize the results. Operated on the Qiskit AER simulator, this class yields results in a dictionary format. It can also emulate an IBMQ backend through the AER simulator, allowing the simulator to use custom quantum gates, basis gates, and coupling maps specific to that backend.
\begin{table}[t]
\centering
\begin{tabular}{|c|c|c|}
\hline
(\( k \)) & 3 Gates (\( \rho_k \)) & 12 Gates (\( \rho_k \)) \\
\hline
  & & \\
$k_0$ & 
$\begin{bmatrix}
0.32+0.i & 0.14-0.08i \\
0.14+0.08i & 0.23+0.i
\end{bmatrix}$ &
$\begin{bmatrix}
0.52+0.i    & 0.35+0.098i \\
0.35-0.09i & 0.269 +0.i 
\end{bmatrix}$ \\
  & & \\
\hline
  & & \\
$k_1$ & 
$\begin{bmatrix}
0.21+0.i & -0.06-0.15i \\
-0.06+0.15i & 0.34+0.i
\end{bmatrix}$ &
$\begin{bmatrix}
0.56+0.i  &  0.13-0.09i \\
0.13+0.08i &0.09+0.i  
\end{bmatrix}$ \\
  & & \\
\hline
  & & \\
$k_2$ & 
$\begin{bmatrix}
0.43+0.i & 0.024-0.04i \\
0.02+0.04i & 0.12+0.i
\end{bmatrix}$ &
$\begin{bmatrix}
0.52+0.i    &    -0.19+0.11i \\
-0.19-0.i & 0.14+0.i 
\end{bmatrix}$ \\
  & & \\
\hline
\end{tabular}

\vspace{1cm} 

\begin{tabular}{|c|c|c|}
\hline
State & Probability (3 Gates) & Probability (12 Gates) \\
\hline
000 & 0.0990 & 0.394637 \\
001 & 0.0553 & 0.065579 \\
010 & 0.2070 & 0.002211 \\
011 & 0.1875 & 0.165900 \\
100 & 0.2479 & 0.267396 \\
101 & 0.0322 & 0.007403 \\
110 & 0.1009 & 0.027074 \\
111 & 0.0701 & 0.069654 \\
\hline
\end{tabular}

\caption{Final results for maximally entangled states with three qubits:
(Top) Reduced density matrices for 3-gate and 12-gate three-qubit quantum circuits; 
(Bottom) Probabilities of quantum states for three-qubit circuits with three and twelve gates.}
\label{table:compact_state_probabilities}
\end{table}
Qiskit, a Python package provided by IBM, is integral to creating and simulating quantum circuits in QUEVO. It offers tools for writing quantum programs and can simulate quantum circuits locally or run them on actual quantum computers via the cloud. To ensure seamless integration and operation of Qiskit within the QUEVO framework, a dedicated Python virtual environment is recommended, preferably set up using Anaconda. 

The \verb!compute_MW_entanglement! method implements the MW measure of entanglement, as a fitness function of the EA. This method processes a state vector representing a quantum circuit's state, converting it into a $2^n$ x $2^n$ tensor, where $n$ is the circuit's qubit count. The entanglement for each qubit $k$ is determined by tracing out all other qubits, summing the squared eigenvalues of their respective reduced density matrices. The results are then normalized by $1 - \frac{1}{n}$, ensuring values fall between 0 and 1. More information and method usage are detailed in the QUEVO1 GitHub repository, accessible at \url{https://github.com/shailendrabhandari/QUEVO1}. Besides the well-known MW measure, we evaluate the entanglement properties of quantum states using the von Neumann entropy as a fitness function. EA creates random density matrices from complex state vectors and calculates the von Neumann entropy based on their eigenvalues. This dual approach in measuring entanglement with MW and von Neumann entropy enhances the quantum circuits' entanglement characteristics, optimized via quantum EAs.

\section*{Supplementary Note II: Evaluation of quantum purity and the MW entanglement measure}
\label{supp4QuantumPurity}

In this appendix, a comprehensive derivation and analysis of the quantum purity of a state is provided, particularly focusing on its relation to the MW measure. 
If the state $\ket{\psi}$ is pure, then its density matrix is given by $\rho = \ket{\psi}\bra{\psi}$, and its purity yields:
\begin{equation}
\operatorname{Tr}[\rho^2] = \operatorname{Tr}[(\ket{\psi}\bra{\psi})(\ket{\psi}\bra{\psi})] 
= \operatorname{Tr}[\ket{\psi}\braket{\psi|\psi}\bra{\psi}] = \operatorname{Tr}[\ket{\psi}\bra{\psi}] = 1.
\end{equation}

The given equation is an expression for the quantum purity of a state $\ket{\psi}$, denoted by $Q(\ket{\psi})$. The purity is a measure of how pure or mixed a quantum state is, and it is defined as the trace of the square of the density matrix $\rho$ that represents the state.
For a pure state, the expression $1 - \frac{1}{n}\sum_{k=0}^{n-1}Tr[\rho_k^2]$ is equal to zero.
For a mixed state, the density matrix can be written as a convex combination of pure states, $\rho = \sum_i p_i \ket{\psi_i}\bra{\psi_i}$, where $p_i$ are probabilities and $\sum_i p_i = 1$. The purity of such a mixed state can not exceed $1$, as one can conclude from: 
\begin{align*}
\operatorname{Tr}[\rho^2] &= \operatorname{Tr} \left[\left(\sum_i p_i \ket{\psi_i}\bra{\psi_i}\right)\left(\sum_j p_j \ket{\psi_j}\bra{\psi_j}\right)\right] \\
&= \sum_i p_i^2 \operatorname{Tr}[\ket{\psi_i}\bra{\psi_i}] + \sum_{i\neq j} p_ip_j \operatorname{Tr}[\ket{\psi_i}\bra{\psi_j}\ket{\psi_j}\bra{\psi_i}] \\
&= \sum_i p_i^2 + \sum_{i\neq j} p_ip_j |\braket{\psi_i|\psi_j}|^2 \\
&\leq \sum_i p_i^2 + \sum_{i\neq j} p_ip_j = \left(\sum_i p_i\right)^2 = 1 \, .
\end{align*}

Therefore, the purity of a mixed state is always less than or equal to one. Using this result, we can see that the expression $1 - \frac{1}{n}\sum_{k=0}^{n-1}Tr[\rho_k^2]$ is a measure of how mixed the state is. If the state is pure, then this expression is zero, and if the state is entangled/mixed, then this expression is positive.

\begin{table}[t]
\centering

\begin{tabular}{|c|c|}
\hline
Qubit (\( k \)) & Reduced Density Matrix (\( \rho_k \))  \\
\hline
& \\
\( k_0 \) & 
$\begin{bmatrix}
0.19027+0.i & -0.079000-0.1031i \\
-0.07902+0.1034i & 0.3575+0.i
\end{bmatrix}$ \\
&\\
\hline
&\\
\( k_1 \) & 
$\begin{bmatrix}
0.26918+0.i & -0.02213+0.04809i \\
-0.02213-0.04859i & 0.23655+0.j
\end{bmatrix}$ \\
&\\
\hline
&\\
\( k_2 \) & 
$\begin{bmatrix}
0.20785+0.i & -0.025037-0.0631i \\
-0.02507+0.06311i & 0.30635+0.i
\end{bmatrix}$ \\
&\\
\hline
&\\
\( k_3 \) & 
$\begin{bmatrix}
0.11746+0.i & -0.01629-0.11099i \\
-0.01649+0.11099i & 0.46981+0.i
\end{bmatrix}$ \\
&\\
\hline

\end{tabular}

\vspace{6mm}
\small{
\begin{tabular}{|c|c|c|c|}
\hline
State & Real Part & Imag. Part & Probability \\
\hline
0000 & -0.11520 & 0.01557 & 0.01351 \\
0001 & -0.18770 & -0.18186 & 0.06830 \\
0010 & 0.09023 & 0.11352 & 0.02103 \\
0011 & 0.37234 & 0.00525 & 0.13866 \\
0100 & 0.17358 & 0.08707 & 0.03771 \\
0101 & -0.02481 & 0.12900 & 0.01726 \\
0110 & 0.14681 & 0.28129 & 0.10068 \\
0111 & -0.08074 & 0.11239 & 0.01915 \\
1000 & -0.11697 & 0.00246 & 0.01369 \\
1001 & 0.22572 & 0.33390 & 0.16244 \\
1010 & -0.07017 & 0.06416 & 0.00904 \\
1011 & -0.10372 & 0.28073 & 0.08957 \\
1100 & -0.19159 & 0.13664 & 0.05538 \\
1101 & 0.03911 & -0.28446 & 0.08245 \\
1110 & 0.24894 & 0.10368 & 0.07272 \\
1111 & -0.26709 & 0.16429 & 0.09833 \\
\hline
\end{tabular}}
\caption{Final outcomes demonstrating maximally entangled states in a four-qubit system: (Top) Reduced density matrices for four-gate quantum circuits; 
(Bottom) Probabilities of quantum states for four-qubit circuits with four gates.}
\label{table:4qubit_combined_probabilities}

\end{table}


\section*{Supplementary Note III: Reduced density matrix as a tool for understanding the purity of a state}
\label{supp3Density_matrix}

The density matrix is a representation of a quantum state that can describe both pure states and mixed states. While the state-vector representation is only suitable for describing pure states, the density matrix can also be used to represent mixed states, which are probabilistic mixtures of pure states. A mixed state can be expressed as a sum of outer products of pure states, each term weighted by a probability. In the density matrix formalism, a mixed state is represented by a Hermitian matrix with non-negative eigenvalues that sum up to one.
The density operator representation is an alternative way to express pure quantum states using a matrix formalism. It is defined as:
\begin{equation} 
\label{dm1}
    \rho \equiv | \psi \rangle \langle \psi |\,, 
\end{equation}

where $|\psi\rangle$ is the state vector representing the pure quantum state. The expression $|\psi\rangle\langle\psi|$ represents the outer product of the state $|\psi\rangle$ with itself, 
\begin{equation}
\rho = 
\begin{bmatrix} \alpha_0 \\ \alpha_1 \\ \vdots \\ \alpha_N \end{bmatrix} \begin{bmatrix} \alpha_0^* & \alpha_1^* & \dots & \alpha_N^* \end{bmatrix} 
=
\begin{bmatrix} |\alpha_0|^2 & \alpha_0 \alpha_1^* & \dots & \alpha_0 \alpha_N^* \\ \alpha_1 \alpha_0^* & |\alpha_1|^2 & \dots & \alpha_1 \alpha_N^* \\ \vdots & \vdots & \ddots & \vdots \\ \alpha_N \alpha_0^* & \alpha_N \alpha_1^* & \dots & |\alpha_N|^2 \end{bmatrix}.
\end{equation}
The density operator has several important properties. It is Hermitian, which means that it is equal to its conjugate transpose. It has a trace equal to one, reflecting the fact that it represents a pure state. Furthermore, it also satisfies the property, $\rho^2 = \rho$, indicating the state is pure. In other words, the density matrix characterizes the pure state.
To illustrate this, let us consider an example of a two-qubit, maximally-entangled pure state $\ket{\psi_{AB}}$, which can be expressed as:
\begin{equation}
     \ket{\psi_{AB}} = \frac{1}{\sqrt{2}} \left ( \ket{00} + \ket{11} \right ) = \frac{1}{\sqrt{2}} \begin{bmatrix} 1 \\ 0 \\ 0 \\ 1 \end{bmatrix}.
\end{equation}

Using the density operator representation, we can express this state as:
\begin{equation}
\rho_{AB}  = 
\left ( \frac{1}{\sqrt{2}} \begin{bmatrix} 1 \\ 0 \\ 0 \\ 1 \end{bmatrix} \right ) \left ( \frac{1}{\sqrt{2}} \begin{bmatrix} 1 & 0 & 0 & 1 \end{bmatrix} \right )  = 
\frac{1}{2} \begin{bmatrix} 1 & 0 & 0 & 1 \\ 0 & 0 & 0 & 0 \\ 0 & 0 & 0 & 0 \\ 1 & 0 & 0 & 1 \\ \end{bmatrix}  \, .
\end{equation}
This matrix satisfies the properties of a density operator, namely, it is Hermitian, positive semi-definite, and has a trace equal to one. The density operator representation provides a compact and convenient way to represent pure states in a matrix formalism, which can be useful for calculations and quantum information processing tasks.

The reduced density matrix $\rho_A$ describes the state of subsystem $A$ after tracing out the degrees of freedom of subsystem $B$. It is obtained by performing a partial trace of the density matrix $\rho_{AB}$ over subsystem $B$ \cite{benenti2007principles}:
\begin{equation}\label{dm2}
    \rho_A = \text{Tr}_B(\rho_{AB}) = \sum_{i=1}^{d_B} \langle i|\rho_{AB}|i\rangle\;, 
\end{equation}
where $d_B$ is the dimension of the Hilbert space of subsystem $B$, and $|i\rangle$ denotes a basis state of subsystem $B$.

The reduced density matrix $\rho_A$ inherits some of the properties of the original density matrix $\rho_{AB}$. For example, it is Hermitian, positive semi-definite, and has a trace equal to one. However, it is generally not pure, even if the original state $\rho_{AB}$ is pure. This is because entanglement between subsystems cannot be removed by taking a partial trace.

$\text{Tr}_B$ in Equation (\ref{dm2}) is an operation known as the partial trace and is used to extract the state of a subsystem of a composite system from its overall density matrix. The partial trace over subsystem $B$ of a tensor product of two operators $|\xi_u\rangle\langle\xi_v|$ and $|\chi_u\rangle\langle\chi_v|$ is given by:
\begin{equation}
    \text{Tr}_B \left (| \xi_u \rangle \langle \xi_v | \otimes | \chi_u \rangle \langle \chi_v | \right ) \equiv | \xi_u \rangle \langle \xi_v | \text{ Tr} \left ( | \chi_u \rangle \langle \chi_v | \right )\;,
\end{equation}
where $| \xi_u \rangle $ and $| \xi_v \rangle $ are arbitrary states in the subspace of $A$, and $| \chi_u \rangle $ and $| \chi_v \rangle $ arbitrary states in the subspace of $B$. $\text{Tr}$ is the standard trace operation, which for two arbitrary states $\text{Tr} \left ( | \chi_u \rangle \langle \chi_v | \right ) = \langle \chi_v |\chi_u \rangle $. Similarly, the reduced density matrix of subsystem $B$ can be obtained by taking the partial trace over subsystem $A$ of the tensor product of two operators:
\begin{equation}
    \text{Tr}_A \left (| \xi_u \rangle \langle \xi_v | \otimes | \chi_u \rangle \langle \chi_v | \right ) \equiv \text{Tr} \left ( | \xi_u \rangle \langle \xi_v | \right ) | \chi_u \rangle \langle \chi_v | .
\end{equation}

As an example, let's reconsider the pure entangled state:\begin{equation}
     | \psi_{AB} \rangle = \frac{1}{\sqrt{2}} \left ( | 0_A 0_B \rangle + | 1_A 1_B \rangle \right ).
\end{equation}

This system is then composed of single-qubit subsystem $A$ with basis vectors $ \left \{ |\xi_1 \rangle, |\xi_2 \rangle \right \} = \{ | 0_A \rangle, | 1_A \rangle \}$, and single-qubit subsystem $B$ with basis vectors $ \left \{ |\chi_1 \rangle, |\chi_2 \rangle \right \} = \{ | 0_B \rangle, | 1_B \rangle \}$. We know that this system is not separable (i.e., $| \chi_{AB} \rangle \neq |\chi_{A}\rangle \otimes |\chi_{B}\rangle$); however, by using the reduced density matrix, we can find a full description for subsystems $A$ and $B$ as follows.

The density matrix of our state $| \psi_{AB} \rangle$ can be expressed in terms of outer products of the basis vectors as
$\rho_{AB} = 
\frac{1}{2} \left [ | 0_A 0_B \rangle \langle 0_A 0_B | + | 0_A 0_B \rangle \langle 1_A 1_B | \right .$
$\left . +| 1_A 1_B \rangle \langle 0_A 0_B | + | 1_A 1_B \rangle \langle 1_A 1_B | \right ] $.
To calculate the reduced density matrix for, let's say, subsystem $B$, we have:
\begin{align*}
    \rho_{B} &= \text{Tr}_A(\rho_{AB}) \\
    &= \frac{1}{2}\left [ \text{Tr}_A(| 0_A 0_B \rangle \langle 0_A 0_B |) + \text{Tr}_A(| 0_A 0_B \rangle \langle 1_A 1_B |) \right .\\
    &\quad + \left . \text{Tr}_A(| 1_A 1_B \rangle \langle 0_A 0_B |) + \text{Tr}_A(| 1_A 1_B \rangle \langle 1_A 1_B |) \right ] \\
    &= \frac{1}{2}\left [ \text{Tr}(| 0_A \rangle \langle 0_A |)| 0_B \rangle \langle 0_B | + \text{Tr}(| 0_A \rangle \langle 1_A |)| 0_B \rangle \langle 1_B | \right .\\
    &\quad + \left . \text{Tr}(| 1_A \rangle \langle 0_A |) | 1_B \rangle \langle 0_B | + \text{Tr}(| 1_A \rangle \langle 1_A |) | 1_B \rangle \langle 1_B | \right ] \\
    &= \frac{1}{2}\left [ \langle 0_A | 0_A \rangle | 0_B \rangle \langle 0_B | + \langle 1_A | 0_A \rangle | 0_B \rangle \langle 1_B | \right .\\
    &\quad + \left . \langle 0_A | 1_A \rangle | 1_B \rangle \langle 0_B | + \langle 1_A | 1_A \rangle | 1_B \rangle \langle 1_B | \right ] \\
    &= \frac{1}{2}\left [ | 0_B \rangle \langle 0_B | + | 1_B \rangle \langle 1_B | \right ]. 
\end{align*}

Starting with an entangled pure state $|\psi_{AB}\rangle$, the reduced density matrix $\rho_B$ of subsystem $B$ emerges as a mixed state. This matrix $\rho_B$ statistically represents subsystem $B$ by averaging out measurements from subsystem $A$. Tracing out $A$ yields $\rho_B$, reflecting the probability distribution of $B$. This approach is crucial for analyzing entangled states in complex systems, allowing insight into individual subsystems within entanglements. Key concepts and details about density matrices are sourced from \cite{benenti2007principles, nielsen_quantum_2002}, especially the section "The density matrix and mixed states" (\url{https://qiskit.org/textbook/ch-quantum-hardware/density-matrix.html##4.-The-Reduced-Density-Matrix--}) 

\section*{Supplementary Note IV: State vector probabilities and density matrices for multi-qubit circuits}
\label{supp2calculatedsv_DM}

In quantum systems, probabilities of states are derived from state vector coefficients within the wave function, typically complex numbers of the form \( a + bi \). The probability \( P \) for a state with coefficient \( c = a + bi \) is given by:
\begin{equation}
    P = |c|^2 = a^2 + b^2,
\end{equation}
where \( a \) represents the real part and \( b \) the imaginary part of the coefficient. Following the computation of reduced density matrices for all qubits, the entanglement of the circuit is assessed using the MW entanglement measure.

This appendix explicitly presents maximally entangled states for three, four, and five qubits. Results are systematically organized in tables, each corresponding to distinct quantum circuit configurations. Table \ref{table:compact_state_probabilities} illustrates the outcomes for a three-qubit system, with reduced density matrices for circuits comprising 3 and 12 gates (Top), and their associated quantum state probabilities (Bottom). Table \ref{table:4qubit_combined_probabilities} explores a four-qubit system, detailing both reduced density matrices and state probabilities for four-gate circuits. Finally, Table \ref{table:5qubit_combined_probabilities} showcases results for a five-qubit system, encompassing reduced density matrices for circuits with five and twelve gates (Top), alongside the state probabilities (Bottom). These tables collectively provide a comprehensive view of quantum state entanglement across varying circuit complexities and gate counts.

\begin{table}[t]
\centering
\footnotesize{
\begin{tabular}{|c|c|c|}
\hline
(\( k \)) & 5 Gates (\( \rho_k \)) & 12 Gates (\( \rho_k \)) \\
\hline
& & \\
\( k_0 \) & 
$\begin{bmatrix}
0.26+0.i & 0.06+0.04i \\
0.06-0.04i & 0.24+0.i
\end{bmatrix}$ &
$\begin{bmatrix}
0.42+0.i & -0.01+0.02i \\
-0.02-0.02i & 0.12+0.i
\end{bmatrix}$ \\
& & \\
\hline
& & \\
\( k_1 \) & 
$\begin{bmatrix}
0.17+0.i & -0.07-0.02i \\
-0.07+0.02i & 0.35+0.i
\end{bmatrix}$ &
$\begin{bmatrix}
0.14+0.i & -0.07-0.02i \\
-0.07+0.025i & 0.40+0.i
\end{bmatrix}$ \\
& & \\
\hline
& & \\
\( k_2 \) & 
$\begin{bmatrix}
0.17+0.i & 0.03+0.02i \\
0.03-0.02i & 0.34+0.i
\end{bmatrix}$ &
$\begin{bmatrix}
0.43+0.i & -0.08-0.04i \\
-0.08+0.04i & 0.13+0.i
\end{bmatrix}$ \\
& & \\
\hline
& & \\
\( k_3 \) & 
$\begin{bmatrix}
0.09+0.i & -0.01+0.04i \\
-0.01-0.04i & 0.47+0.i
\end{bmatrix}$ &
$\begin{bmatrix}
0.27+0.i & -0.07+0.08i \\
-0.06-0.08i & 0.24+0.i
\end{bmatrix}$ \\
& & \\
\hline
& & \\
\( k_4 \) & 
$\begin{bmatrix}
0.23+0.i & -0.13-0.09i \\
-0.13+0.09i & 0.32+0.i
\end{bmatrix}$ &
$\begin{bmatrix}
0.38+0.i & -0.01+0.01i \\
-0.01-0.012i & 0.14+0.i
\end{bmatrix}$ \\
& & \\
\hline 
\end{tabular}}

\vspace{6mm}

\small{
\begin{tabular}{|c|c|c||c|c|c|}

\hline
State & Prob. 5 Gates & Prob. 12 Gates & State & Prob. 5 Gates & Prob. 12 Gates \\
\hline
00000 & 0.00951 & 0.00397 & 10000 & 0.00030 & 0.02975 \\
00001 & 0.02023 & 0.04046 & 10001 & 0.00382 & 0.01760 \\
00010 & 0.01381 & 0.06053 & 10010 & 0.00464 & 0.02113 \\
00011 & 0.02924 & 0.04221 & 10011 & 0.07066 & 0.02118 \\
00100 & 0.01295 & 0.05451 & 10100 & 0.00216 & 0.00176 \\
00101 & 0.01313 & 0.00470 & 10101 & 0.03519 & 0.00101 \\
00110 & 0.02475 & 0.00025 & 10110 & 0.05050 & 0.03929 \\
00111 & 0.04909 & 0.02533 & 10111 & 0.06792 & 0.00579 \\
01000 & 0.01619 & 0.19855 & 11000 & 0.06648 & 0.00968 \\
01001 & 0.02338 & 0.01998 & 11001 & 0.00260 & 0.01616 \\
01010 & 0.01261 & 0.02199 & 11010 & 0.01035 & 0.06754 \\
01011 & 0.05647 & 0.07690 & 11011 & 0.07818 & 0.00342 \\
01100 & 0.01677 & 0.06047 & 11100 & 0.01954 & 0.00957 \\
01101 & 0.00801 & 0.00635 & 11101 & 0.06265 & 0.03323 \\
01110 & 0.17413 & 0.03243 & 11110 & 0.01923 & 0.00157 \\
01111 & 0.02745 & 0.00311 & 11111 & 0.00029 & 0.06759 \\
\hline
\end{tabular}}
\caption{Final outcomes demonstrating maximally entangled states in a five-qubit system: (Top) Reduced density matrices for five-gate and twelve-gate three-qubit quantum circuits; 
(Bottom) Probabilities of quantum states for five qubit circuits with five and twelve gates.}
\label{table:5qubit_combined_probabilities}
\end{table}

\end{document}